\documentclass[12pt]{article}
\usepackage{epsfig}

\def\nn{\nonumber}
\def\im{\mbox{ Im }}
\def\etl{{\em et al.}}

\begin{document}

\title{QCD SUM RULES DESCRIPTION OF NUCLEONS IN
ASYMMETRIC NUCLEAR MATTER
}
\author{
E.G. Drukarev, M.G. Ryskin, V.A. Sadovnikova\\
Petersburg Nuclear Physics Institute, \\
Gatchina, St.~Petersburg 188300,
Russia}
\date{}
\maketitle

\begin{abstract}
We calculate the nucleon parameters in  isospin asymmetric nuclear
matter using the QCD sum rules.  The nucleon self-energies $\Sigma_v$
and $\Sigma^*_s$ are expressed in terms of the in-medium values
of QCD condensates.  The simple approximate expressions for the
 self-energies are obtained in terms of these condensates.
Relation between successive inclusion of the condensates and the
meson-exchange picture of the nucleon interaction with medium is
analyzed. The values of the self-energies and of the symmetry
energy agree with those obtained by the methods of nuclear
physics.

\end{abstract}

\begin{center}
{\bf I. INTRODUCTION }
\end{center}

In this paper we investigate the vector and scalar self-energies of
nucleons in nuclear matter composed by the neutrons and protons,
distributed with densities $\rho_n$ and $\rho_p$. We calculate
the dependence on the total density
\begin{equation}\label{1a}
\rho\ =\ \rho_p+\rho_n
\end{equation}
and on the asymmetry parameter
\begin{equation}\label{2a}
\beta\ =\ \frac{\rho_n-\rho_p}{\rho_p+\rho_n}
\end{equation}
(another conventional presentation of the asymmetry parameter is
$\beta=\frac{N-Z}{N+Z} = 1 - \frac{2Z}A$ with $N$ and $Z$ standing for
the total number of neutrons and protons, while $A=N+Z$). We present
also the equations for the single-particle potential energies. The
results are obtained by QCD sum rules approach.

The QCD sum rules were invented by Shifman {\em et al.} \cite{1} to
express the hadron parameters through the vacuum expectation values of
QCD operators. Being initially used for the mesons, the method was
expanded by Ioffe \cite{2,3} to the description of the baryons. The
approach succeeded in describing the static characteristics as well as
some of the dynamical characteristics of the hadrons in vacuum --- see,
e.g., the reviews \cite{4,5}.

The main idea is to consider the correlation function
\begin{equation}
\label{3}
\Pi_0(q)\ =\ q_\mu\gamma^\mu\Pi^q_0(q^2)+I\Pi^I_0(q^2)\ ,
\end{equation}
($I$ stands for the unit $4\times4$ matrix) describing
the propagation of the system with the quantum numbers of the hadron, in
the different regions of values of the momentum $q^2$, where certain
information on its behavior is available. The asymptotic freedom of QCD
enables to present $\Pi^j_0(q^2)$ ($j=q,I$) at $q^2\to-\infty$ as the
power series of $q^{-2}$ and QCD coupling $\alpha_s$. On the other
hand, the imaginary part of $\Pi_0(q^2)$ at $q^2>0$ can be described in
terms of the observable hadrons. This prompts to consider the
dispersion relation for the function $\Pi^j_0(q^2)$ \cite{1}
\begin{equation}\label{4}
\Pi_0(q^2)\ =\ \frac1\pi\int\frac{\im\Pi_0(k^2)}{k^2-q^2}\ dk^2
\end{equation}
at $q^2\to-\infty$. The coefficients of the expansion of the left-hand
side (lhs) of the functions $\Pi^j_0(q^2)$ in powers of $q^{-2}$ are the
expectation values of the local operators constructed of the quark and
gluon fields, which are called "condensates". Such presentation, known
as the operator product expansion (OPE) \cite{6} provides the
perturbative expansion of the short-distance effects, while the
nonperturbative physics is contained in the condensates. The usual
treatment of the right-hand side (rhs) of Eq.~(\ref{4}) consists in
"pole + continuum" presentation, in which the lowest laying pole is
singled out while the higher states are approximated by the continuum
described by the pure perturbative OPE contribution.
Thus, Eq.~(\ref{4}) ties the values of QCD condensates with the
characteristics of the lowest hadronic state. Such interpretation
requires that the contribution of the pole to the rhs of Eq.~(\ref{4})
exceeds the contribution of the continuum.

The OPE of the lhs of Eq.~(\ref{4}) becomes increasingly true, while the
value of $|q^2|$ increases. On the other hand, the "pole + continuum"
model becomes more accurate while $|q^2|$ decreases. The important
assumption is that the two presentations are close in certain
intermediate region of the values of $q^2$. To improve the overlap of the
QCD and the phenomenological descriptions, one usually applies certain
mathematical anzats,  i.e., the Borel transform. The Borel transformed
dispersion relations (\ref{4}) are known as QCD sum rules \cite{1,2}.

For example, the QCD sum rules for the nucleon provided a connection
between the nucleon mass and the scalar quark condensate $\langle0|\bar
qq|0\rangle$ \cite{2,3}. A more detailed analysis  requires inclusion
of the expectation values of the higher dimension, e.g., the gluon
condensate $\langle0|\frac{\alpha_s}\pi G_{\mu\nu}G_\mu^\nu|0\rangle$
and the four-quark condensate $\langle0|\bar qq\bar qq|0\rangle$.  Two
other unknowns of nucleon QCD sum rules equations are the
residue of the nucleon pole $\lambda^2_0$ and the effective
continuum threshold $W^2_0$.

Later the QCD sum rules were applied for investigation of modified
nucleon parameters in the symmetric nuclear matter \cite{7,8}.
They were based on the Borel-transformed dispersion relation for the
function $\Pi_m(q)$ describing the propagation of the system with the
quantum numbers of the nucleon (the proton) in the nuclear matter.
Considering nuclear matter as a system of $A$ nucleons with momenta
$p_i$, one introduces the vector
\begin{equation} \label{5}
p\ =\ \frac{\Sigma p_i}A\ ,
\end{equation}
which is thus $p\approx(m,0)$ in the rest frame of the
matter. The function $\Pi_m(q)$ can be presented as
$\Pi_m(q)=\Pi_m(q^2,\varphi(p,q))$ with the arbitrary function
$\varphi(p,q)$ being kept constant in the dispersion relations in
$q^2$.

The general form of the function $\Pi_m$ can thus be presented as
\begin{equation} \label{7}
\Pi_m(q)\ =\ q_\mu\gamma^\mu\Pi^q_m(q^2,s)+p_\mu\gamma^\mu\Pi^p_m
(q^2,s)+I\Pi^I_m(q^2,s)\ .
\end{equation}
The in-medium QCD sum rules are the Borel-transformed dispersion
relations for the components $\Pi^j_m(q^2,s)$ $(j=q,p,I)$
\begin{equation} \label{8}
\Pi^j_m(q^2,s)\ =\ \frac1\pi\int\frac{\im\Pi^j_m(k^2,s)}{k^2-q^2}\
dk^2\ .
\end{equation}
The spectrum of the function $\Pi_m(q)$ is much more complicated than
that of the function $\Pi_0(q)$. The choice of the function
$\varphi(p,q)$ is dictated by the attempts to separate the singularities
connected with the nucleon in the matter from those connected with the
properties of the matter itself. Since the latter manifest themselves as
singularities in the variable $s=(p+q)^2$, the separation can be done by
putting $\varphi(p,q)=(p+q)^2$ and by fixing \cite{7,8,9}
\begin{equation} \label{6}
\varphi(p,q)\ =\ (p+q)^2\ \equiv\ s\ =\ 4E_{0F}^2
\end{equation}
with $E_{0F}$ being the total nucleon energy at
the Fermi surface.

By using Eq.~(\ref{8}) the characteristics of the nucleon in
nuclear matter can be expressed through the in-medium values of
QCD condensate. The possibility of extension of  "pole + continuum"
model \cite{1,2} to the case of finite densities was shown in
\cite{7}--\cite{10}.

The lowest order of OPE of the lhs of Eq.~(\ref{8}) can be presented
in terms of the vector and scalar condensates \cite{7}--\cite{10}.
Vector condensates $v^i_\mu=\langle M|\bar q\,^i\gamma_\mu
q^i|M\rangle$ of the quarks with the flavor $i$ ($|M\rangle$ denotes
the ground state of the matter) are the linear functions of the nucleon
densities $\rho_n$ and $\rho_p$. In the asymmetric matter both SU(2)
symmetric and asymmetric condensates
$$
v_\mu=\langle M|\bar u(0)\gamma_\mu
u(0)+\bar d(0)\gamma_\mu d(0)|M\rangle = v^u_\mu+v^d_\mu
$$ and
$$
v^{(-)}_\mu
=\langle M|\bar u(0)\gamma_\mu u(0)-\bar d(0)\gamma_\mu
d(0)|M\rangle=v^u_\mu-v^d_\mu
$$
obtain nonzero values.  In the rest frame
of the matter $v^i_\mu =v^i\delta_{\mu0}$, $v_\mu=v\delta_{\mu0}$,
$v^{(-)}_\mu=v^{(-)}\delta_{\mu0}$. We can present
\begin{equation}
\label{9}
v^i\ =\ \langle n|\bar q^i\gamma_0 q^i|n\rangle\rho_n
+\langle p|\bar q^i\gamma_0 q^i|p\rangle\rho_p.
\end{equation}
The values
$\langle N|\bar q^i\gamma_0 q^i|N\rangle$
are just the numbers of the
valence quarks in the nucleons
$\langle n|\bar u\gamma_0 u|n\rangle=
\langle p|\bar d\gamma_0 d|p\rangle=1$,
$\langle p|\bar u\gamma_0 u|p\rangle=
\langle n|\bar d\gamma_0 d|n\rangle=2$,
and
thus
\begin{equation}
\label{10}
v^u=\ \rho_n+2\rho_p=\ \rho\left(\frac32-\frac\beta2\right), \quad
v^d=\ 2\rho_n+\rho_p=\ \rho\left(\frac32+\frac\beta2\right).
\end{equation}
Hence, we obtain
\begin{equation}
\label{11}
v(\rho)\ =\ v_N\rho\ , \qquad v^{(-)}(\rho,\beta)\ =\
\beta v^{(-)}_N\rho \end{equation}
with
\begin{equation}
\label{12}
v_N\ =\ 3\ , \qquad v^{(-)}_N\ =\ -1\ .
\end{equation}

The lhs of Eq.~(\ref{8}) contains the SU(2) symmetric scalar condensate
\begin{equation}
\label{13}
\kappa_m(\rho)\ =\ \langle M|\bar u(0)u(0)+\bar d(0)d(0)|M\rangle\ ,
\end{equation}
and SU(2) asymmetric one
\begin{equation}
\label{14}
\zeta_m(\rho,\beta)\ =\ \langle M|\bar u(0)u(0)-\bar d(0)d(0)|M\rangle\ .
\end{equation}
These condensates can be presented as
$$
 \kappa_m(\rho)=\kappa_0+\kappa(\rho)\ ,
$$
$\kappa_0=\kappa_m(0)$ is the vacuum value,
\begin{equation}\label{15}
 \kappa(\rho)=\kappa_N\rho+\ldots\ , \quad \kappa_N=\langle N|\bar
uu+\bar dd|N\rangle\ ,
\end{equation}
and
\begin{equation}
\zeta_m(\rho,\beta)\ =\ -\beta(\zeta_N\rho+\ldots)\ , \quad \zeta_N=\langle
p|\bar uu-\bar dd|p\rangle\ .
\label{16}
\end{equation}
The dots in the rhs of
Eqs.~(\ref{15}) and (\ref{16}) denote the terms, which are nonlinear in
$\rho$. In the gas approximation such terms should be omitted. The
SU(2) invariance of vacuum was assumed in Eq.~ (\ref{16}). The
expectation value $\kappa_N$ is related to the $\pi N$ sigma term
$\sigma_{\pi N}$, i.e., \cite{11}
\begin{equation}
\kappa_N\ =\ \frac{2\sigma_{\pi
N}}{m_u+m_d}
\label{17}
\end{equation}
with $m_{u,d}$ standing for the current
masses of the light quarks, while the value of $\sigma_{\pi N}$ can be
extracted from experimental data on low energy $\pi N$ scattering
\cite{12,13}. However, the value $\zeta_N$ should be calculated under
certain model assumptions on the quark structure of the nucleon.
These were the condensates of dimension $d=3$.

Turning to the condensates of dimension $d=4$, we find for the gluon
condensate
$$
 g_m(\rho)\ =\ \langle M|\frac{\alpha_s}\pi\ G^2(0)|M\rangle\ =\
g_0+g(\rho)\ ,
$$
$g_0 = g_m(0)$ is the vacuum value,
\begin{equation}
   g(\rho)\ =\ g_N\rho+\ \ldots
\label{18}
\end{equation}
with the nucleon expectation value
\begin{equation}
g_N\ =\ \langle N|\frac{\alpha_s}\pi\ G^2|N\rangle\ \approx\
-\frac89\,m\ ,
\label{19}
\end{equation}
obtained in \cite{14} in a model-independent way.

Also, the nonlocal condensate $\langle N|\bar q(0)\gamma_\mu q(x)|N
\rangle$ provides the contributions of $d=4$ and those of the higher
dimension. The term of the dimension $d=4$ is
\begin{equation}
\label{20}
\langle N|\bar q\,^i(0)\gamma_\mu D_\nu q^i(0)|N\rangle\ =\ \left(
g_{\mu\nu}-\frac{4p_\mu p_\nu}{m^2}\right)mx_2
\end{equation}
with $x_2$ standing for the second moment of the nucleon structure
function \cite{9}.

The condensates of dimensions $d=3,4$ determine the leading OPE terms
of the order $q^2\ln q^2$ and $\ln q^2$ of the lhs of Eq.~(\ref{8}).
The next to leading terms of the order $1/q^2$ are determined by the
expectation values of the four-quark operators. The importance of these
contributions was analyzed in \cite{15}. In the gas approximation they
can be presented  in terms of the nucleon expectation values. The
latter can be obtained in framework of certain models.

We shall analyze the sum rules in the gas approximation. It is a
reasonable starting point, since the nonlinear contributions to the
most important scalar condensate $\kappa(\rho)$ are relatively small at
the densities of the order of the phenomenological saturation density
$\rho_0=0.17\,\rm fm^{-3}$  of the symmetric matter \cite{10}. The QCD
sum rules approach, applied to the description of the nucleon
self-energies in the symmetric matter \cite{16} provided the results
which are consistent with those, obtained by the methods of nuclear
physics.

In the case of symmetric matter $(\beta=0)$, the leading OPE terms
$(d=3,4)$ can be either calculated or expressed in terms of the
observables. In the asymmetric case we need a model for the calculation
of the expectation value $\zeta_N$, defined by Eq.~(\ref{16}). In most
of the quark models the nucleon is treated as a system of the valence
quarks and the isospin symmetric sea of quark--antiquark pairs. Under
this assumption the condensate $\zeta_N$ is determined by the valence
quarks. In the models with the nonrelativistic valence quarks
$\zeta_N=1$. In more realistic relativistic models $\zeta_N<1$ due to
the relativistic reduction.

The calculations of the four-quark condensates require model
assumptions on the structure of the nucleon. The complete set of the
four-quark condensates was obtained in \cite{17} by using the
perturbative chiral quark model (PCQM). The chiral quark model,
originally suggested in \cite{18}, was developed further in
\cite{19}--\cite{21}. In the PCQM the nucleon is treated as a system
of relativistic valence quarks moving in an effective static field. The
valence quarks are supplemented by a perturbative cloud of
pseudoscalar mesons, in agreement with the requirements of the chiral
symmetry.
This model reflects the main points of the nowadays view on the
structure of nucleon. The latter is treated as the system of the
valence quark and the sea-quarks, contained in pions. The
chiral properties are respected. In \cite{17} the SU(2) version
of PCQM, which includes only the pions, have been used.  Thus,
in the calculations, which include the terms of the order
$1/q^2$ of OPE obtained in framework of PCQM
we must use the PCQM value $\zeta_N=0.54$
\cite{19}.

In the rhs of the sum rules we describe the nucleon by the relativistic
in-medium propagator \cite{22}
\begin{equation}
G^{-1}_N\ =\ q_\mu\gamma^\mu-m - \Sigma
\label{21}
\end{equation}
with the total self-energy
\begin{equation}
\Sigma\ =\ q_\mu\gamma^\mu\Sigma_q+p_\mu\gamma^\mu\,\frac{\Sigma_p}m+
\Sigma_s\ .
\label{22}
\end{equation}
We shall use the QCD sum rules for the calculation of the nucleon
characteristics
\begin{equation}
\Sigma_v=\frac{\Sigma_p}{1-\Sigma_q}\ , \quad
m^*=\frac{m+\Sigma_s}{1-\Sigma_q}\ , \quad \Sigma^*_s=m^*-m\ ,
\label{23}
\end{equation}
identified with the vector self-energy,  Dirac effective mass
and the effective scalar self-energy --- see, e.g., \cite{22}. Two other
parameters, to be determined in the approach are the in-medium shifts
of the effective values of the nucleon residue $\delta\lambda^2_m$ and
of the continuum threshold $\delta W^2_m$. We present also the result
for the single-particle potential energies
\begin{equation}
U\ =\ \Sigma^*_s+\Sigma_v\ .
\label{24}
\end{equation}
We trace the dependence of these characteristics on the total
density $\rho$ and on the asymmetry parameter $\beta$ --- Eqs.
(\ref{1a}) and (\ref{2a}).

Our approach includes only the strong interactions between the
nucleons. The electromagnetic and weak interactions are neglected.

In the asymmetric matter the characteristics obtain different values
for the proton and neutron. Considering the asymmetry parameter in the
interval $-1\le\beta\le1$, one can see that the values of any parameter
$P_N$ for the proton $(p)$ and neutron $(n)$ are connected as
\begin{equation}
P_n(\beta)\ =\ P_p(-\beta)\ .
\label{25}
\end{equation}
We shall present all the values for the proton, using Eq.~(\ref{25}) to
obtain those for the neutron.

We carry out the calculations in the gas approximation, including only
terms linear in $\rho$ in the lhs of the sum rules. Thus we can neglect
the Fermi motion of the nucleons of the matter, which manifest
  themselves in higher order of density, and put
\begin{equation}
s\ =\ 4m^2
\label{26}
\end{equation}
in Eq.~(\ref{6}) .  Having in mind future extensions of the
approach, we shall keep the dependence on $s$, using Eq.
(\ref{26})  for the specific computations.

In the simplest Hartree approximation (without multiparticle
forces) $\Sigma_p$ is linear in density.  If one neglects the Fermi
motion of the nucleon, as we did, the same is true for the scalar
self-energy $\Sigma_s$.  Thus, the parameters $\Sigma_v$ and $m^*$
(\ref{23}) exhibit nonlinear behavior due to nonzero value of
$\Sigma_q$ ($\Sigma_q=0$ in the mean field approximation). In similar
way, the characteristics $\Sigma_v$ and $m^*$ are not linear in $\rho$
and $\beta$ in our approach due to the relativity large lhs of the sum
rules for the structure $\Pi^q_m$.  This reflects distribution of the
baryon charge between the pole and continuum. Another reason of the
nonlinear effects (although, numerically less important) is the
nonlinear structure of the sum rules equations.

The in-medium nucleon QCD sum rules
present nucleon interactions with matter in terms of
the exchange by uncorrelated quark-antiquark
pairs.
The latter exchanges correspond to the meson exchanges with the
same quantum numbers. Thus the contributions of the lowest order
OPE terms can be viewed to the exchange by the isoscalar and
isovector vector and scalar mesons, with the point-like vertices
of the interactions, having the standard Lorentz structure.
Inclusion of the nonlocal effects in the quark condensates
corresponds to inclusion of nonlocal effects in the vertices of
the interactions between the nucleons of the matter and the
mesons. The four-quark condensates, in which two of the quark
operators are averaged over the vacuum state correspond to the
anomalous Lorentz structures of the vertices of the
nucleon-meson interactions.  In terms of hadronic degrees of
freedom the four-quarks condensates, in which all the quark
operators act inside the nucleons correspond to the exchanges by
the strongly correlated four-quark systems. Such exchanges can
be viewed, e.g.,  as those by two mesons, interacting with the
nucleon at the same point.  This is illustrated by Figs.~1 and 2.

We compare our results for the nucleon self-energies and potential
energies with those, obtained by the traditional methods of nuclear
physics. We find a promising consistency of the results.

We recall in brief the main points of QCD sum rules approach in
vacuum and in nuclear matter in Sec.~II. We present the general
equations in Sec.~III, providing the solutions in Sec.~IV. We compare
the results to those, obtained by the methods of nuclear physics in
Sec.~V.  We summarize in Sec.~VI.

\begin{center}
{\bf I. GENERAL EQUATIONS}
\end{center}
\begin{center}
{\bf A. Sum rules in vacuum}
\end{center}

To make the paper self-consistent, we recall the main points of the QCD
sum approach in vacuum \cite{1,2}. The function $\Pi_0(q^2)$ (often
referred to as "polarization operator") is presented as
\begin{equation}
\Pi_0(q)\ =\ i\int d^4xe^{i(qx)}\langle0|Tj(x)\bar j(0)|0\rangle
\label{27}
\end{equation}
with $j$ being the three-quark local operator (often referred to as
"current") with the proton quantum numbers. The usual choice is \cite{2}
\begin{equation}
j(x)\ =\ \varepsilon^{abc}\left[u^{aT}(x)C\gamma_\mu u^b(x)\right]
\gamma_5\gamma_\mu d^c(x)\ ,
\label{28}
\end{equation}
where $T$ denotes a transpose and $C$ is the charge  conjugation
matrix. The upper indices denote the colors.

The lhs of Eq.~(\ref{4}) is approximated by several lowest terms of
OPE, i.e., $\Pi^j_0(q^2)\approx\Pi^{j,OPE}_0(q^2)$. The empirical data
are used for the spectral function $\im\Pi_0(q^2)$ on the rhs of Eq.
(\ref{4}).  Namely, it is known, that the lowest laying state is the
bound state of three quarks, which manifests itself as a pole in the
(unknown) point $k^2=m^2$. Since the next singularity is the branching
point $k^2=W^2_{ph}=(m+m_\pi)^2$, one can present
\begin{equation}
\im\Pi^j_0(k^2)\ =\ \lambda^2_N a_j\delta(k^2-m^2)+f^j(k^2)\theta
(k^2-W^2_{ph})
\label{29}
\end{equation}
with $\lambda^2_N$ being the residue at the nucleon pole. Here $j=q,I$
correspond to the two Lorentz structures of Eq.~(\ref{3}), $a_q=1$,
$a_I=m$. Thus, Eq.~(\ref{4}) takes the form
\begin{equation}
\Pi^{j,OPE}_0(q^2)\ =\ \frac{\lambda^2_Na_j}{m^2-q^2}+\frac1\pi
\int\limits^\infty_{W^2_{ph}} \frac{f^j(k^2)}{k^2-q^2}\ dk^2\ .
\label{30}
\end{equation}
Of course, the detailed structure of the spectral density $f^j(k^2)$
cannot be resolved in such approach. The further approximations are
based on the asymptotic behavior
\begin{equation}
\label{31}
f^j(k^2)\ =\ \frac1{2i}\ \Delta\Pi^{j,OPE}_0(k^2)
\end{equation}
at $k^2\gg|q^2|$ with $\Delta$ denoting the discontinuity. The
discontinuity is caused by the logarithmic contributions of the
perturbative OPE terms. The usual anzats consists in extrapolation of
Eq.~(\ref{31}) to all the values of $k^2$, replacing also the physical
threshold $W^2_{ph}$ by the unknown effective threshold $W^2_0$, i.e.,
\begin{equation}
\frac1\pi\int\limits^\infty_{W^2_{ph}}\frac{f^j(k^2)}{k^2-q^2}\ dk^2\ =\
\frac1{2\pi i} \int\limits^\infty_{W^2_0}
\frac{\Delta\Pi^{j,OPE}_0(k^2)}{k^2-q^2}\ dk^2\ .
\label{32}
\end{equation}
Thus Eq.~(\ref{4}) takes the form
\begin{equation}
\Pi^{j,OPE}_0(q^2)\ =\ \frac{\lambda^2_Na_j}{m^2-q^2} +
\frac1{2\pi i}\int\limits^\infty_{W^2_0}\frac{\Delta\Pi^{j,OPE}_0
(k^2)}{k^2-q^2}\ dk^2\ .
\label{33}
\end{equation}

The lhs of Eq.~(\ref{33}) contains QCD condensates. The rhs of Eq.~(\ref{33})
contains three unknown parameters: $m,\lambda^2_N$ and $W^2_0$. The OPE
becomes increasingly true while the value $|q^2|$ increases. The "pole +
continuum" model is more accurate at the smaller values of $|q^2|$.
Thus one can expect Eq.~(\ref{33}) to be true in a certain limited interval
of the values of $|q^2|$. To improve the overlap of OPE and
phenomenological descriptions one usually applies the Borel transform
defined as
\begin{eqnarray}
&& Bf(q^2)\ =\ \lim\limits_{Q^2,n\to\infty}\frac{(Q^2)^{n+1}}{n\,!}
\left(-\frac d{dQ^2}\right)^nf(q^2)\ \equiv\ \tilde f(M^2)\ ,
\nonumber\\
\label{34}
&& Q^2\ =\ -q^2, \quad M^2\ =\ Q^2/n
\end{eqnarray}
with $M$ called the Borel mass. It is important in the applications to
the sum rules that Borel transform eliminates the polynomials and
emphasizes the contribution of the lowest state in rhs of Eq.~(\ref{33})
due to relation
\begin{equation}
B\ \frac1{m^2-q^2}\ =\ e^{-m^2/M^2}\ .
\label{35}
\end{equation}
The Borel-transformed form of Eq.~(\ref{33}) reads
\begin{equation}
\widetilde\Pi^{j,OPE}_0(M^2)\ =\ \lambda^2_Ne^{-m^2/M^2}a_j
+\frac1{2\pi i}\int\limits^\infty_{W^2_0}dk^2e^{-k^2/M^2}\Delta\,
\Pi^{j,OPE}_0 (k^2)
\label{36}
\end{equation}
and is known as QCD sum rules. Actually, there are two sum rules for
the structures $\Pi^q_0$ and $\Pi^I_0$ of the function
$\Pi_0(q)$ defined by Eq.~(\ref{3}).

It appeared to be more convenient to work with Eq.~(\ref{36})
multiplied by the numerical factor $32\pi^4$. The two sum rules for the
nucleon in vacuum can be presented in the form \cite{2}
\begin{eqnarray}
\label{37}
&& L^q_0(M^2,W^2_0)\ =\ \Lambda_0(M^2)\ , \\
&& L^I_0(M^2,W^2_0)\ =\ m\Lambda_0(M^2)
\label{38}
\end{eqnarray}
with
\begin{equation}
\Lambda_0(M^2)\ =\ \lambda^2_0\, e^{-m^2/M^2}\ .
\label{39}
\end{equation}
Here $\lambda^2_0=32\pi^4\lambda^2_N$,
\begin{eqnarray}
&& L^q_0(M^2,W^2_0)=32\pi^4\bigg(\widetilde\Pi_0\,^{q,OPE}(M^2) -
\frac1{2\pi i}
\int\limits^\infty_{W^2_0}dk^2e^{-k^2/M^2}\Delta\,
\Pi^{q,OPE}_0(k^2)\bigg), \nonumber\\
&& L^I_0(M^2,W^2_0)\ =\ 32\pi^4\bigg(\widetilde\Pi_0^{I,OPE}(M^2) -
\frac1{2\pi i} \int\limits^\infty_{W^2_0}dk^2
e^{-k^2/M^2}\Delta\, \Pi^{I,OPE}_0(k^2)\bigg).
\label{40}
\end{eqnarray}

The lhs of Eqs.~(\ref{37}) and (\ref{38}) \cite{2,3} have been obtained
with the account of the condensates of dimension $d=8$, i.e., with the
account of the terms of the order $1/q^4$ in OPE of the functions
$\Pi^{j,OPE}_0$
\begin{eqnarray}
\label{41}
L^q_0(M^2,W^2_0) &=& \frac{M^6E_2}{L^{4/9}}+ \frac{bE_0 M^2}{4L^{4/9}}
+\frac43 a^2L^{4/9} - \frac13 \frac{\mu^2_0}{M^2}a^2,\\
L^I_0(M^2,W^2_0) &=& 2aM^4E_1
-\frac{ab}{12}+\frac{272}{81} \frac1{M^2} \frac{\alpha_s}\pi a^3
\label{42}
\end{eqnarray}
with the traditional notations
$a=-(2\pi)^2\langle0|\bar qq|0\rangle=-2\pi^2\kappa_0$ (we assumed the
isotopic invariance of vacuum $\langle0|\bar
uu|0\rangle=\langle0|\bar dd|0\rangle=\langle0|\bar
qq|0\rangle$), $b=(2\pi)^2g_0$, $\mu^2_0=0.8\,\rm GeV^2$. Here
$E_i$ are the functions of the ratio $W_0^2/M^2$:
$E_i=E_i(W_0^2/M^2)$. They are determined by the formulas

\begin{equation}
E_0(x)=1-e^{-x}, \quad E_1(x) = 1-(1+x)e^{-x},\quad
 E_2(x) = 1-\left(1+x +\frac{x^2}2\right)e^{-x} .
\label{43}
\end{equation}
The factor
\begin{equation}
L(M^2) \ =\ \frac{\ln M^2/\Lambda^2}{\ln\nu^2/\Lambda^2}
\label{44}
\end{equation}
accounts for the anomalous dimension, i.e.,
the most important corrections of the order $\alpha_s$ enhanced by the
"large logarithms". In Eq.~(\ref{44}) $\Lambda=\Lambda_{QCD}=0.15\,$GeV,
while $\nu=0.5\,$GeV is the normalization point of the characteristic
involved. Note that the two last terms on the rhs of Eq.~(\ref{41})
originate from the four-quark condensates $\langle0|\bar u\Gamma^X
u\bar u\Gamma^X u|0\rangle$ and can be expressed through the single
term $(\langle0|\bar qq|0\rangle)^2$ only in framework of factorization
hypothesis \cite{1}--\cite{3}. Also, the last term on the rhs of Eq.
(\ref{42}) is the six-quark condensate, evaluated in the same
approximation.

The matching of the lhs and rhs of Eqs.~(\ref{37}) and (\ref{38}) have
been achieved \cite{2,3} in the domain
\begin{equation}
0.8\mbox{ GeV}^2\ <\ M^2\ <\ 1.4\,\rm GeV^2
\label{45}
\end{equation}
providing the values of the vacuum parameters
\begin{equation}
\lambda^2_0=1.9\mbox{ GeV}^6\ , \quad W^2_0=2.2\mbox{ GeV}^2,
\label{46}
\end{equation}
if $m=0.94$ GeV.

\begin{center}
{\bf B. Sum rules in nuclear matter}
\end{center}

The OPE terms of the polarization operator in nuclear matter
\begin{equation}
\Pi_m(q)\ =\ i\int d^4 xe^{i(qx)}\ \langle M|Tj(x)\bar j(0)|M\rangle
\label{47}
\end{equation}
contains the in-medium values of QCD condensates. Some of these
condensates vanish in vacuum, obtaining non-zero values only in medium.
The other ones just change their values compared to the vacuum ones.

The spectrum of the function $\Pi_m(q)$ is much more complicated, than
that of the vacuum function $\Pi_0(q^2)$. However, \cite{7}--\cite{10},
the spectrum of the function $\Pi_m(q^2,s)$ at fixed value of $s$ can
be described by the "pole + continuum" model at least until we include
the terms of the order $\rho^2$ in the OPE of $\Pi_m(q^2,s)$.

Using Eqs.~(\ref{21}) - (\ref{23}), we present the nucleon propagator
as
\begin{equation}
G_N\ =\ Z_N\cdot\frac{q_\mu\gamma^\mu-p_\mu\gamma^\mu
(\Sigma_v/m)+m^*}{q^2-m^2_m}
\label{48}
\end{equation}
with $\Sigma_v$ and $m^*$ defined by Eq.~(\ref{23}), while
the new position of the nucleon pole is
\begin{equation}
m^2_m\ =\ \frac{(s-m^2)\Sigma_v/m-\,\Sigma^2_v+m^{*2}}{1+\Sigma_v/m}\ ,
\label{49}
\end{equation}
and
\begin{equation}
Z_N\ =\ \frac1{(1-\Sigma_q)(1+\Sigma_v/m)}\ .
\label{50}
\end{equation}

Thus, we shall present the dispersion relations for the functions
$\Pi^j_m(q^2,s)$ $(j=q,p,I)$  in the form
\begin{equation}
\Pi^{j,OPE}_m(q^2,s)=\frac{Z_N\lambda^2_m b_j}{m^2_m-q^2}+\frac1{2\pi i}
\int\limits^\infty_{W^2_m} \frac{\Delta_{k^2}\Pi^{j,OPE}(k^2,s)}{
k^2-q^2}
\label{51}
\end{equation}
with $b_q=1$, $b_p=-\Sigma_v$, $b_I=m^*$. The Borel-transformed
sum rules take the form
\begin{eqnarray}
\label{52}
L^q_m(M^2,W^2_m) &=& \Lambda_m(M^2)\ ,\\
\label{53}
L^p_m(M^2,W^2_m) &=& -\Sigma_v\Lambda_m(M^2)\ ,\\
\label{54}
L^I_m(M^2,W^2_m) &=& m^*\Lambda_m(M^2)
\end{eqnarray}
with
\begin{equation} \label{55}
\Lambda_m(M^2)\ =\ \lambda^{*2}_m e^{-m^2_m/M^2}\ .
\end{equation}
Here
\begin{equation} \label{56}
\lambda^{*2}_m\ =\ \lambda^2_m\cdot Z_N
\end{equation}
is the effective value of the nucleon residue in nuclear matter.

Actually, we shall solve the sum rules equations, subtracting the
vacuum effects
\begin{eqnarray} \label{57}
&& L^q(M^2,W^2_m,W^2_0)\ =\ \Lambda_m(M^2)-\Lambda_0(M^2)\ ,
\\
\label{58}
&& L^p(M^2,W^2_m)\ =\ -\Sigma_v\Lambda_m(M^2)\ ,
\\
\label{59}
&& L^s(M^2,W^2_m,W^2_0)=m^*\Lambda_m(M^2)-m\Lambda_0(M^2)
\end{eqnarray}
with $L^j=L^j_m-L^j_0$. The functions $\Lambda_0$, $L^q_0$ and $L^I_0$
are defined by Eqs.~(\ref{39}), (\ref{41}) and (\ref{42}), while
$L^p_0=0$.

\begin{center}
{\bf III. CONTRIBUTIONS TO THE LEFT-HAND SIDE OF THE SUM RULES}
\end{center}

We shall use the current, presented by Eq.~(\ref{28}) for the
calculation of the function $\Pi_m(q)$, defined by
Eq.~(\ref{47}). The possibility of using it in isospin asymmetric
system requires additional argumentation, which is given in
Appendix A.

We shall include the terms of the order $q^2\ln q^2$, $\ln q^2$ and
$1/q^2$ in the lhs of the sum rules (\ref{51}). This corresponds to the
terms of the order $M^4$, $M^2$ and $\bf 1$ in the lhs of the Borel
transformed equations (\ref{57})--(\ref{59}). We shall include
subsequently the contributions of three types. The terms
$\ell^j_m(M^2)$ stand for the lowest order local condensates. They
contribute to $q^2\ln q^2$ and $\ln q^2$ terms of Eq.~(\ref{51}). These
contributions are similar to simple exchanges by isovector
vector and scalar mesons between the nucleon on the rhs of
Eq.~(\ref{51}) and the nucleons of the matter, Figs.~1(a),2(a). The
terms $u^j_m(M^2)$ are caused by the nonlocalities of the vector
condensate, Fig.~1(d), contributing to the terms of the order $\ln q^2$
and $1/q^2$.  They correspond to the account of the form factors in the
vertices between the isovector mesons couple to the nucleons,
Fig.~2(d). Finally, $\omega^j(M^2)$ describes the contributions of the
four-quark condensates, Figs.~1(b),1(c), being of the order $1/q^2$.
They
correspond to the two-meson exchanges (or to exchanges by
four-quark mesons, if there are any) and to somewhat more
complicated structure of the meson-nucleon vertices, Fig.~2(c).
Thus we present the lhs of Eqs.  (\ref{52})--(\ref{54}) as
\begin{equation} \label{60}
L^j_m\ =\ \ell^j_m+u^j_m+w^j_m
\end{equation}
and the lhs of Eqs.  (\ref{57})--(\ref{59})
\begin{equation} \label{61}
L^j\ =\ \ell^j+u^j+w^j
\end{equation}
with $\ell^j=\ell^j_m-\ell^j_0$, $u^j=u^j_m$,
$\omega^j=\omega^j_m-\omega^j_0$, while $\ell^j_0$ and $\omega^j_0$ are
the corresponding contributions in the vacuum case.

The calculation of the function $\Pi_m(q^2,s)$ defined by Eq.
(\ref{47}) is based on the presentation of the single-quark propagator
in medium
\begin{equation}
\langle M|Tq^i_\alpha(x)\bar q\,^i_\beta(0)|M\rangle =
G_{\alpha\beta}(x)-\frac14 \langle M|\bar q\,^i(0)\gamma_\mu
q^i(x)|M\rangle\gamma^\mu_{\alpha\beta}-\frac14\langle M|
\bar q\,^i(0)q^i(x)|M\rangle \delta_{\alpha\beta}
\label{62}
\end{equation}
with $G(x)=(ix_\mu\gamma^\mu)/(2\pi^2x^4)$ being the free
propagator of the quark in the chiral limit.
In the OPE calculations we neglect the values of
the current masses of the light quarks since they are small
in our scale.
Recall that $i$
denotes the light quark flavor.  In the lowest orders of OPE two
 of the quarks are described by the free propagators and only
one of the quarks is presented by the second or the third term
of the rhs of Eq.~(\ref{62}). Inclusion of the higher order OPE
terms will require the two-quark in-medium propagators.

At $x=0$ the matrix elements in the second and third terms on the rhs
are just the vector and scalar condensates defined by Eqs.~(\ref{9})
and (\ref{15}).  The  contribution of the bilocal configurations can be
expressed in terms of the higher moments and twists of the nucleon
structure functions \cite{9}.

The bilocal operators on the rhs of Eq.~(\ref{62}) are not gauge
invariant.  The gauge  invariant expression, achieved by substitution
\cite{23}
\begin{equation}
q^i(x)\ =\ q^i(0)+x_\alpha D_\alpha q^i(0)+\frac12\,x_\alpha x_\beta
D_\alpha D_\beta q^i(0)+ \cdots
\label{63}
\end{equation}
with $D_\alpha$ standing for the covariant derivatives, provides the
infinite set of the local condensates. The expectation values in
Eq.~(\ref{62}) depend on the variables $(px)$ and $x^2$. In the gas
approximation we need only the nucleon matrix elements. For the vector
structure the general form is \begin{equation} \label{64}
\theta^i_\mu(x)=\langle M|\bar q^i(0)\gamma_\mu q^i(x)|M\rangle =
\frac{p_\mu}m\,\Phi^i_a ((px),x^2)+ix_\mu m\Phi^i_b((px),x^2)
\end{equation} with $q^i(x)$ defined by Eq.~(\ref{63}).

The contribution of the vector structure (\ref{64}) to the polarization
operator $\Pi_m(q)$ can be presented as
\begin{equation}
\label{65}
\Pi^V_m(q)=\frac{4i}{\pi^4}\int\frac{d^4x}{x^8}\left(x^2\hat\theta^u
+(x,\theta^u+\theta^d)\hat x\right)e^{i(qx)}.
\end{equation}
Here we denoted
\begin{equation}
\hat a\ =\ a_\mu\gamma^\mu
\label{66}
\end{equation}
for any 4-vector $a_\mu$. For the scalar structure
\begin{equation}
\label{67}
\Pi^I_m(q)\ =\ -\frac{2i}{\pi^4}\int\frac{d^4x}{x^6}\,\langle
M|\bar d(0)d(x)|M\rangle e^{i(qx)}.
\end{equation}

\begin{center}
{\bf A. Local condensates of the lowest dimensions}
\end{center}

To include the local vector and scalar quark condensates of the lowest
dimensions, we must put $x=0$ in Eq.~(\ref{64}) and in the matrix
element $\langle M|\bar d(0)d(x)|M\rangle$ in  Eq.~(\ref{67}). Thus
$$
\Phi^i_a(0,0)=
\langle n|\bar q^i\gamma_0 q^i|n\rangle\rho_n
+\langle p|\bar q^i\gamma_0 q^i|p\rangle\rho_p ,
$$  while
$$\langle M|\bar
d(0)d(0)|M\rangle=\langle0|\bar dd|0\rangle+\rho_p\langle p|\bar dd|p
\rangle+\rho_u\langle n|\bar dd|n\rangle .
$$
We must add also the
contribution of the gluon condensate. Technically it can be obtained by
replacing the vacuum gluon condensate by it's in-medium value --- Eqs.
(\ref{18}) and (\ref{19}).

The contributions of these local condensates can be expressed as
\begin{eqnarray}
\label{68}
&& \ell^q\ =\ f^q_v(M^2,W^2_m)v^q(\rho)+f^q_g(M^2,W^2_m)g(\rho),
\\
&& \ell^p\ =\ f^p_v(M^2,W^2_m)v^p(\rho,\beta)\ ,
\label{69}\\
&& \ell^I\ =\ f^I_\kappa (M^2,W^2_m)t^I(\rho,\beta)\ ,
\label{70}
\end{eqnarray}
with the dependence on $\rho$ and $\beta$ being contained in the
factors
\begin{equation}
\label{71}
v^q(\rho)=3\rho, \quad v^p(\rho,\beta)=3\rho\left(1-\frac\beta4\right),
\quad t^I(\rho,\beta)=\rho(\kappa_N+\zeta_N\beta),
\end{equation}
and the function $g(\rho)$, given by Eqs.~(\ref{18}) and (\ref{19}).
The other functions are \cite{16}
\begin{eqnarray}
&&
f^q_v(M^2,W^2_m)
=-\frac{8\pi^2}3\frac{(s-m^2)M^2E_{0m}-M^4E_{1m}}{m\,L^{4/9}},
\nn \\
&& f^q_g(M^2,W^2_m)=\frac{\pi^2 M^2E_{0m}}{L^{4/9}},
\nn \\
&& f^p_v(M^2,W^2_m)=-\frac{8\pi^2}3\ \frac{4M^4\,E_{1m}}{L^{4/9}}\ ,
\nn\\
&& f^I_\kappa(M^2,W^2_m)=-4\pi^2M^4E_{1m}\ .
\label{72}
\end{eqnarray}
The notation $E_{km}\,(k=0,1)$, Eq.~(\ref{43}), means that the
functions depend on the ratio $W^2_m/M^2$.

Now we must find the $\beta$-dependence of the nucleon self-energies
and also of the parameters $W^2_m$ and $\lambda^{*2}_m$. Note that
there  is a simple solution of Eqs.~(\ref{52}) and (\ref{54})
\begin{eqnarray}
&& \Sigma_v(\rho,\beta)\ =\
\Sigma_v(\rho,0)\left(1-\frac\beta4\right)\ ,
\label{73}\\
&& m^*(\rho,\beta;\kappa_N,\zeta_N)\ =\
m^*(\rho,0;\kappa_N+\beta\zeta_N,0)\ , \label{74}\\ &&
W^2_m(\rho,\beta)\ =\ W^2_m(\rho,0)\ , \label{75}
\end{eqnarray}
which is true with the same accuracy as the solutions for the symmetric
matter $(\beta=0)$ \cite{16}. Indeed, assuming that $W^2_m$ does not
change with $\beta$ (\ref{73}), we find that the function
$\Lambda_m(M^2)$ in the lhs of Eq.~(\ref{52}) should not depend on
$\beta$~\footnote{
The numerical solution, which will be considered in Sec. IV
provides indeed $W^2_m(\rho,\beta)\approx W^2_m(\rho,0)$ with
the accuracy of $10\%$.}.
This leads to Eqs.~(\ref{74}) and (\ref{75}).

\begin{center}
{\bf B. Inclusion of the nonlocal condensates}
\end{center}

Now we present the functions in the rhs of Eq.~(\ref{64}) as
\begin{equation}
\label{76}
\Phi^i_{a(b)}\ =\ \rho_n\phi^i_{n,a(b)}+\rho_p\phi^i_{p,a(b)}\ .
\end{equation}
Due to SU(2) invariance they can be presented in terms of the proton
functions $\phi^i_{a(b)}=\phi^i_{p,a(b)}$
\begin{equation}
\label{77}
\Phi^i_{a(b)}\ =\ \rho_n\phi^{i'}_{a(b)}+\rho_p\phi^i_{a(b)}
\end{equation}
with $i=u,d,\  i'\neq i$.

Expansion in powers of $x^2$ corresponds to the expansion of the
function $\Pi_m(q)$ in powers of $q^2$. To obtain the terms of the
order $q^{-2}$ it is sufficient to include two lowest terms of the
expansions in powers of $x^2$. One can present \cite{9,24}
\begin{equation}
\label{78}
\phi^i_{a(b)}((px),x^2)\ =\ \int\limits^1_0 d\alpha
e^{-i\alpha(px)}f^i_{a(b)}(\alpha,x^2)
\end{equation}
with
\begin{equation}\label{79}
f^i_{a(b)}(\alpha,x^2)=\eta^i_{a(b)}(\alpha) + \frac18
x^2m^2\xi^i_{a(b)}(\alpha) .
\end{equation}
Here $\eta^i_a(\alpha)=f^i_a(\alpha,0)$ is
the contribution of the quarks with the flavor $i$ to the asymptotic
of the nucleon structure function $\eta(\alpha)=\eta^u_a(\alpha)
+\eta^d_a(\alpha)$, normalized by the condition
\begin{equation} \label{80}
\int^1_0 d\alpha\eta(\alpha)\ = \ 3
\end{equation}
with the rhs presenting just the number of the valence quarks in the
nucleon. Thus, expansion of the function
$$
\varphi^i_a(px)=\phi^i_a((px),0)
$$
in powers of $(px)$ is expressed
through the moments of the distributions $\eta^i_a(\alpha)$. The
moments are well known --- at least, those, which are numerically
important. Also, the first moments
$$
\xi^i_a\ =\ \int\limits^1_0\xi^i_a(\alpha)d\alpha
$$
of the distribution $\xi^i_a(\alpha)$ were calculated in
\cite{25} by the QCD sum rules method:
\begin{equation}
\label{81}
\xi^u\ =\ -0.24 , \quad \xi^d\ =\ -0.09 , \quad
\xi = \xi^u + \xi^d = -0.33 .
\end{equation}
The moments
of the function $\eta^i_b(\alpha)$ can be obtained by using the
equations of motion $D_\alpha\gamma^\alpha q^i(x)=m_iq^i(x)$. Thus, in
the chiral limit \cite{9}
\begin{eqnarray}
&& \langle\varphi^i_b\rangle =\frac14\langle\varphi^i_a\alpha\rangle\ ,
\nonumber\\
&& \langle\varphi^i_b\alpha\rangle=\frac15\left(\langle\varphi^i_a
\alpha^2\rangle -\frac14\langle\xi^i\rangle \right),
\nonumber\\
&&  \langle \xi^i_b\rangle\ =\ \frac16\langle \xi^i_a\alpha\rangle\ .
\label{82}
\end{eqnarray}
Here we denoted
\begin{equation}\label{83}
\langle f\rangle=\int^1_0 d\alpha f(\alpha)
\end{equation}
for any function $f(\alpha)$.

Note that the nonlocality of the scalar condensate, i.e., of the last
term on the rhs of Eq.~(\ref{62}) does not manifest itself in the terms
up to $1/q^2$. The first derivative in $(px)$, as well all the
derivatives of the odd order vanish in the chiral limit due to QCD
equation of motion. The next to leading order of the expansion in
powers of $x^2$ vanishes due to certain cancellations \cite{8} as well
as in the case of vacuum \cite{2} for the particular choice of the
operator $j(x)$ presented by Eq.~(\ref{28}). We neglect the terms,
containing more complicated gluon condensates, since they are
numerically small \cite{26}.

Using Eqs.~(\ref{64}) and (\ref{77}) we obtain for the vector structure
(\ref{65})
\begin{eqnarray}
\label{84}
&& \Pi^V_m(q)\ =\ \Pi^{Va}_m(q)+\Pi^{Vb}_m(q)\ ,
\\
&& \Pi^{Va}_m(q)\ =\
\frac{2i}{\pi^4 m}\left\{\int\frac{d^4x}{x^8}\left(\hat
px^2+2\hat x(px)\right)(\varphi^u_a+\varphi^d_u)e^{i(qx)}\rho
\right.  \nn\\ &&\left. -\ \beta\hat
p\int\frac{d^4x}{x^6}(\varphi^u_a-\varphi^d_a)
e^{i(qx)}\rho\right\} ,
\nn\\
&&\Pi^{Vb}_m(q)=-\frac{6 m}{\pi^4}\int\frac{d^4x}{x^6}\hat
x(\varphi^u_b+ \varphi^d_b)e^{i(qx)}\rho
+\frac{2\beta}{\pi^4}\int\frac{d^4x}{x^6} \hat
x(\varphi^u_b-\varphi^d_b)e^{i(qx)}.  \nn \end{eqnarray}

The further calculations, employing Eqs.~(\ref{78}) and (\ref{79}) are
similar to those, carried out for the case of symmetric matter
\cite{16} --- see Appendix B.

Finally, the higher moments and higher twists of the nucleon
structure functions provide the contributions $u^i$ to the lhs
$L^i_m$ of the sum rules --- Eq.~(\ref{60})
\begin{eqnarray}
&&
u^q(M^2)\ =\ (u^q_{N,1}(M^2) + \beta u^q_{N,2}(M^2))\rho\ ,
\nn \\
&& \quad
u^q_{N,1}(M^2)=\frac{8\pi^2}{3L^{4/9}m}\left[
-\frac52 m^2 M^2 E_{0m} \langle\eta\alpha\rangle
+\frac32 m^2 (s-m^2)\langle\xi\rangle\right]\ ,
\nn\\
&&\quad  u^q_{N,2}(M^2)=\frac{8\pi^2}{3L^{4/9}m}\left[
\frac32\,m^2M^2E_{0m}(\langle\eta^u\alpha\rangle
-\langle\eta^d\alpha\rangle) \right] ;
\label{85}
\\
\nn\\
\nn\\
&&  u^p(M^2)\ =\ (u^p_{N,1}(M^2) + \beta u^p_{N,2}(M^2))\rho\ ,
\nn\\
&& \quad
u^p_{N,1}(M^2)=\frac{8\pi^2}{3L^{4/9}}\bigg[-5\left(M^4 E_{1m}
-(s-m^2)M^2 E_{0m}\right)\langle\eta\alpha\rangle
\nn\\
&&
\quad -\ \frac{12}5 m^2M^2E_{0m}\langle\eta\alpha^2\rangle
+\ \frac{18}5m^2M^2E_{0m}\langle\xi\rangle\bigg]\ ,
\nn \\
&& \quad
u^p_{N,2}(M^2)=\frac{8\pi^2}{3L^{4/9}}\bigg[
 3\left(M^4E_{1m}-(s-m^2)M^2E_{0m}\right)
\langle(\eta^u-\eta^d)\alpha\rangle
\nn\\
\label{86m}
&&
\quad+\ \frac95\,m^2M^2E_{0m}\langle(\eta^u-\eta^d)\alpha^2\rangle
-\frac{27}{10}\,m^2M^2E_{0m}
\langle(\xi^u-\xi^d)\rangle \bigg] ;
\\
\nn\\
\nn\\
&&
u^I(M^2)\ =\ 0\ .
\label{85a}
\end{eqnarray}
Here we denote $L=L(M^2)$, $E_{1m}=E_1(W^2_m/M^2)$,
$E_{0m}=E_0(W^2_m/M^2)$, Eq.~(\ref{43}), the value of $\xi$ is
determined by Eq.~(\ref{81}).

\begin{center}
{\bf C. Inclusion of the four-quark condensates}
\end{center}

The exchange by two quark-antiquark pairs between the current
(\ref{28}) and the matter is described in terms of the four-quark
expectation values
\begin{equation} \label{86}
H^{XY}_m(\rho) = \langle M|\bar u\Gamma^Xu\bar u\Gamma^Yu|M
\rangle\ \mbox{ and }\    R^{XY}_m(\rho)=\langle M|\bar
d\Gamma^Xd\bar u \Gamma^Yu|M\rangle
\end{equation}
with $\Gamma^{X,Y}$ being the basic $4\times4$ matrices
\begin{equation}
\label{87}
\Gamma^I=I, \quad \Gamma^{Ps}=\gamma_5, \quad \Gamma^V=\gamma_\mu,
\quad \Gamma^A=\gamma_\mu\gamma_5, \quad \Gamma^T=\frac
i2(\gamma_\mu\gamma_\nu-\gamma_\nu\gamma_\mu),
\end{equation}
acting on the Lorentz indices of the quark operators. We did not
display the color indices in Eq.~(\ref{86}), keeping in mind that the
quark operators are color antisymmetric.

In the gas approximation
\begin{eqnarray}
&& H^{XY}_m(\rho,\beta)\ =\ H^{XY}_m(0)+\rho_n h^{XY}_n+\rho_ph^{XY}_p
, \nn\\
&& R^{XY}_m(\rho,\beta)\ =\ R^{XY}_m(0)+\rho_n r^{XY}_n+\rho_pr^{XY}_p.
\label{88}
\end{eqnarray}

The characteristics $h_N^{XY}$ and $r_N^{XY}$ can be presented as
\begin{eqnarray}
h^{XY}_N &=&\frac56\left(\langle0|\bar u\Gamma^Xu|0\rangle\langle N|
\bar u\Gamma^Yu|N\rangle+\langle0|\bar u\Gamma^Yu|0\rangle\langle
N|\bar u\Gamma^Xu|N\rangle\right)
\nn\\
\label{89}
&+& \langle N|(\bar u\Gamma^Xu\cdot\bar u\Gamma^Yu)_{int}|N\rangle\ ,
\\
r^{XY}_N &=&\frac23\left(\langle0|\bar d\Gamma^Xd|0\rangle\langle
N|\bar u\Gamma^Yu|N\rangle+\langle0|\bar u\Gamma^Yu|0\rangle\langle
 N|\bar d\Gamma^Xd|N\rangle\right)
\nn \\
&+& \langle N|(\bar d\Gamma^Xd\cdot \bar u\Gamma^Yu)_{int}|N\rangle\ .
\label{90}
\end{eqnarray}
The first two ("factorized") terms in the rhs of Eqs.~(\ref{89})
and (\ref{90}) describe two quark operators, acting on the
vacuum state, while the other two operators act inside the
nucleon. The last terms describe the "internal" action of all
four operators inside the nucleon. This is shown by the lower
index $"int"$.  The coefficients 5/6 and 2/3 on rhs of Eqs.
(\ref{89}) and (\ref{90}) present the weights of the
color-antisymmetric states \cite{16}.

The contribution of the four-quark expectation values to the in-medium
change of the polarization operator can be written as
\begin{eqnarray}
&& (\Pi)_{4q}\ =\ (\Pi_m)_{4q}-(\Pi_0)_{4q}=\ \frac{\rho_n}{q^2}\left(
\sum_{X,Y}\mu_{XY}h^{XY}_n+\sum_{X,Y}\tau_{XY}r^{XY}_n\right)
\nn \\
&& +\ \frac{\rho_p}{q^2}\left(\sum_{X,Y}\mu_{XY}h^{XY}_p+\sum_{X,Y}
\tau_{XY}r^{XY}_p\right).
\label{91}
\end{eqnarray}
The matrices $\mu_{XY}$ and $\tau_{XY}$ can be obtained by using the
general expression for the function $\Pi_m(q)$ presented in \cite{26}
\begin{eqnarray}
\mu_{XY} &=& \frac{\theta_Y}{16}\mbox{ Tr }(\gamma_\alpha
\Gamma^X\gamma_\beta\Gamma^Y)\gamma_5\gamma^\alpha\hat
q\gamma^\beta\gamma_5\ ,
\nn \\
\tau_{XY} &=& \frac{\theta_Y}4\mbox{ Tr }(\gamma_\alpha\hat q
\gamma_\beta \Gamma^Y)\gamma_5\gamma^\alpha
\Gamma^X\gamma^\beta\gamma_5\ , \quad
\hat q\ =\ q_\mu\gamma^\mu\ .
\label{92}
\end{eqnarray}
Here $\theta_Y=1$ if $\Gamma^Y$ has a vector
or tensor structure, while $\theta_Y=-1$ in the scalar, pseudoscalar
and axial cases. The sign is determined  by that of the commutator
between matrix $\Gamma^Y$ and the charge conjugation matrix $C$ ---
Eq.~(16).

The products $\mu_{XY}h_i^{XY}$ obtain nonzero values if the matrices
$\Gamma^X$ and $\Gamma^Y$ have the same Lorentz structure. In this case
all the structures presented by Eq.~(\ref{87}) contribute to
$(\Pi_m)_{4q}$. The products $\tau_{XY}r^{XY}$  do not turn to zero
only if $\Gamma^Y$ has a vector or axial structure. In the latter case
$\Gamma^X$ should be an axial matrix as well. In the former case
$\Gamma^X$ can be either Lorentz scalar or Lorentz vector.

We denote $h_i^{XX}=h_i^X$, $\mu_{XX}=\mu_X$ and $r_i^{XX}=r_i^X$,
$\tau_{XX}=\tau_X$ for the similar Lorentz structures $X$ and $Y$. The
scalar and pseudoscalar expectation values are Lorentz scalars. Thus,
their contributions can be expressed through single parameters
$a^S$, $a^{Ps}$, $a^{SV}$. The
latter is true also for the scalar-vector expectation value $r^{SV}$.
We obtain
\begin{equation}
\label{93}
\mu_S=-\frac {\hat q}2\ , \quad \mu_{Ps}=\frac{\hat q}2\ , \quad
(\tau_{SV})_{\mu}=-2q_{\mu}\ .
\end{equation}
In the other channels the four-quark condensates have more
complicated structure. In the vector and axial channels
\begin{eqnarray}
&& \left(h_i^{V(A)}\right)_{\mu\nu}\
=\ a^{V(A)}_{h,i}g_{\mu\nu}+b^{V(A)}_{h,i}\frac{p_\mu
p_\nu}{m^2}\ ,
\nn \\
\label{94}
&& \left(r_i^{V(A)}\right)_{\mu\nu}\
=\ a^{V(A)}_{r,i}g_{\mu\nu}+b^{V(A)}_{r,i}\frac{p_\mu
p_\nu}{m^2}\ .
\end{eqnarray}

In the tensor channel
\begin{equation}
\label{95}
\left(h^T_i\right)_{\mu\nu,\rho\tau}\
=\ a^T_{h,i}s_{\mu\nu,\rho\tau}+b^T_{h,i}t_{\mu\nu,\rho\tau}
\end{equation}
with
\begin{eqnarray}
s_{\mu\nu,\rho\tau} &=& g_{\mu\rho}g_{\nu\tau}-g_{\mu\tau}g_{\nu\rho}\ ,
\nn \\
t_{\mu\nu,\rho\tau} &=& \frac1{m^2}\bigg(p_\mu p_\rho g_{\nu\tau} +
p_\nu p_\tau g_{\mu\rho}-p_\mu p_\tau g_{\nu\rho}-p_\nu p_\rho
g_{\mu\tau} \bigg).
\label{96}
\end{eqnarray}

Using Eq.~(\ref{92}) we obtain
\begin{eqnarray}
&& \mu_Vh_i^V=-a^V_{h,i}\hat q-b^V_{h,i}\,\frac{\hat p(pq)}{m^2}\ ,
\quad \mu_Ah_i^A=a^A_{h,i}\hat q+b_{h,i}^A\,\frac{\hat
p(pq)}{m^2}\ , \nn\\
&&\tau_{Vi}r_i^V=\left(-10a^V_{r,i}-2b^V_{r,i}\right)\hat q-2b_{r,i}^V
\, \frac{\hat p(pq)}{m^2}\ ,
\nn \\
&& \tau_{Ai}r^A_i=\left(-6a^A_{r,i}-2b^A_{r,i}\right)\hat q+2b_{r,i}^A
\frac{\hat p(pq)}{m^2}\ ,
\nn \\
\label{97}
&& \mu_T h_i^T = b^T_{h,i}\left(-\frac{\hat q}2+\frac{2\hat
p(pq)}{m^2}\right).
\end{eqnarray}
We can write Eq.~(\ref{91}) in the form
\begin{equation}
\label{98}
(\Pi)_{4q}=\ B^q\frac{\hat q}{q^2}+B^p\frac{(pq)}{m^2}\frac{\hat
p}{q^2}+B^I\frac{mI}{q^2}
\end{equation}
with each of the coefficients $B^j$ being the composition of the
contributions of the neutrons and protons of the matter $B^j_n$ and
$B^j_p$, i.e.,
\begin{equation}
\label{99}
B^j\ =\ B^j_n\rho_n+B^j_p\rho_p\ .
\end{equation}
Using Eqs.~(\ref{93}) -- (\ref{95}) we obtain
\begin{eqnarray}
&& B^q_i=\left(-\frac12 a^S_{h,i}+\frac12 a^{Ps}_{h,i}-a^V_{h,i}
+a^A_{h,i}-\frac12b^T_{h,i}\right)-2\left(5a^V_{r,i}+b^V_{r,i}+
3a^A_{r,i}+b^A_{r,i}\right),
\nn\\
&& B^p_i\ =\ -b^V_{h,i}+b^A_{h,i}+2b^T_{h,i}-2(b^V_{r,i}-b^A_{r,i}),
\nn\\
&& B^I_i\ \approx\ -2a^{SV}_i\,\frac{(pq)}{m^2}\ .
\label{100}
\end{eqnarray}
Here we denoted $a^S_i=h^{SS}_i$, $a^{Ps}_i=h_i^{PsPs}$. Recall that the
lower index $"i"$ stands for the neutron or the proton. In the
last equality we defined $r^{SV}_i = a^{SV}_ip_\mu/m$,
 neglecting the nonlocal contributions to the scalar-vector
condensate.  However in the scalar channel such terms provide
the contributions of the order $1/q^2$ to the rhs of Eq.
(\ref{98}). They will be included below --- Appendix C.

The calculations of the contributions to rhs of Eq.~(\ref{100}) require
the model assumption on the structure of the nucleon. The complete set
of the four-quark condensate was obtained in \cite{17} by using the
perturbative chiral quark model (PCQM). The chiral quark model,
originally suggested in \cite{18}, was developed further in
\cite{19}--\cite{21}. In the PCQM the nucleon is treated as a system of
relativistic valence quarks moving in an effective static field. The
valence quarks are supplemented by a perturbative cloud of the
pseudoscalar mesons in agreement with the requirements of the chiral
symmetry. In \cite{17} the simplest SU(2) version of PCQM, which
includes only the pions, have been used. The parameters which
enter Eq.~(\ref{100}) are calculated in this version of PCQM.

There are three types of contributions to the four-quark condensate in
the framework of this approach.  All four operators can act on the
valence quarks.  Also, four operators can act on the pion. There
is also a possibility that two of the operators act on the
valence quarks while the other two act on the pions. Following
\cite{17} we speak of the "interference terms" in the latter case.

To obtain the contribution of the pion cloud, we need the expectation
values of the four-quark operators in pions. The latter have been
deduced in \cite{27} by using the current algebra technique. It was
shown in \cite{16}, that in the case of the symmetric matter there is a
remarkable cancellation of the pion contributions in the function
$\Pi_m(q)$. Since the pion contents of the asymmetric matter manifests
itself in the different intensities of $\pi^+$ and $\pi^-$ fields,
while the four-quark expectation values in $\pi^+$ and $\pi^-$ mesons
are the same, the cancellation takes place in this case as well. This
cancellation takes place in any model of the nucleon which treats the
pion cloud perturbatively. Thus, the contributions of the four-quark
condensates come from the terms, determined by the valence quarks only
and from the interference terms.

This enables us to present Eq.~(\ref{89}) as
\begin{equation}
\label{101}
h_i^X=2\cdot\frac56\langle 0|\bar u\Gamma^Xu|0\rangle\langle N_i|(\bar
u\Gamma^Xu)_v|N_i\rangle +\langle N_i|(\bar u\Gamma^Xu\bar u\Gamma^Xu)_1
|N_i\rangle\ .
\end{equation}
Here the lower index $"v"$ means that the operators act on the valence
quarks only. The lower index $"1"$ corresponds to the sum of the term
in which all the four operators act on the valence quarks and the term
in which two of the operators act on the valence quarks while the other
two act on pions. Of course, the first ("factorized") term in
the rhs of Eq.~ (\ref{101}) obtains a nonvanishing value only in
the scalar case $\Gamma^X=I$.

The expectation values of the operators of different flavors, providing
nonvanishing contributions to the rhs of Eq.~(\ref{90}) are the
scalar-vector condensate
\begin{equation}
\label{102}
r^{SV}_{i\mu}=2\cdot\frac23\langle0|\bar dd|0\rangle\langle N_i|\bar
u\gamma_\mu u|N_i\rangle+\langle N_i|(\bar dd\bar u\gamma_\mu
u)_{int}|N_i\rangle
\end{equation}
and
\begin{equation} \label{103}
r^X_{i\mu\nu}\ =\ \langle N_i|\left(\bar d\Gamma^X_\mu d\bar
u\Gamma^X_\nu u\right)_1|N_i\rangle
\end{equation}
with $X$ standing for vector or axial structures. The meaning of the
lower index $"1"$ is the same as in Eq.~(\ref{101}).

Note that among the interference terms contributing to the four-quark
condensates, there is so-called "vertex
interference", in which one of the PCQM vertices of the self-energy of
the valence quark is replaced by the four-quark operator. Some of such
terms contain the matrix elements $\langle0|\bar
u\gamma_5d|\pi^-\rangle$ and $\langle0|\bar d\gamma_5u|\pi^+\rangle$,
contributing to the expectation values $\langle N|\bar u\gamma_5d\bar
d\gamma_5u|N\rangle$, being connected with the matrix elements
$\langle N|\bar d\Gamma^Xd\bar u\Gamma^Xu|N\rangle$ of all structures
$\Gamma^X$ by the Fierz transform. On the other hand, they depend on
the values of the quark masses, since $\langle0|\bar
u\gamma_5d|\pi^-\rangle=-\frac{i\sqrt2\,F_\pi M^2_\pi}{m_u+m_d}$ with
$M_\pi$ and $F_\pi$  denoting the mass and the decay constant of
pion.  In a somewhat straightforward approach one substitutes
the current quark masses.  Following more sophisticated models
of the pions \cite{28} one should substitute the constituent
quark masses, thus obtaining the values, which are negligibly
small in our scale.

Using the complete set of the nucleon four-quark expectation values
\cite{17}, we obtain
\begin{equation}\label{104}
(\Pi)_{4q}\ =\ \left(A^q_{4q}(\beta)\frac{\hat q}{q^2}+A^p_{4q}(\beta)
\frac{(pq)}{m^2}\frac{\hat p}{q^2}+A^I_{4q}(\beta)m\frac I{q^2}
\right)\frac a{(2\pi)^2}\rho
\end{equation}
with the coefficients $A^i_{4q}(\beta)$ being determined by Eqs.
(\ref{101})--(\ref{103}), while
\begin{equation}\label{105}
a\ =\ -(2\pi)^2\langle 0|\bar uu|0\rangle\ .
\end{equation}
We use the value $\langle 0|\bar uu|0\rangle$=(-241 MeV)$^3$,
corresponding to $a$=0.55 GeV$^3$, employed in \cite{3}. Note that $a$
is just a convenient scale for presentation of the results. It does not
reflect the chiral properties of $\Pi_{4q}$.

The $\hat q$ term results mainly as the sum of the expectation value of
the product of the four $u$-quark operators, described by the first
("factorized") term
on the rhs of Eq.~(\ref{101}), and that of the product of two $u$
and two $d$-quark operators in the vector channel --- Eq.~(\ref{103}).
The former contributions depend on $\beta$, while the latter do not.
Thus, the coefficient $A^q_{4q}$ depends on $\beta$ strongly.
The $\hat p$ term is determined mostly by the expectation value
(\ref{103}) in the vector channel, with the protons and neutrons
providing the equal contributions.
This explains the weak dependence of the parameter $A^p_{4q}$ on
$\beta$. The coefficient $A^I_{4q}$ is dominated by the first term on
the rhs of Eq.~(\ref{102}), providing stronger dependence on $\beta$.
The calculations give
\begin{equation}
\label{106}
A^q_{4q}=-0.11-0.21\beta, \quad A^p_{4q}=-0.57+0.09\beta, \quad
A^I_{4q}=1.90-0.92\beta\ .
\end{equation}

In the simplified model of the pion, which does not include the
renormalization of the quark masses by the interactions, the value of
the coefficient $A^q_{4q}$ is somewhat different
\begin{equation}
\label{107}
A^q_{4q}\ =\ 0.25-0.22\,\beta\ ,
\end{equation}
while the values of $A^p_{4q}$ and $A^I_{4q}$ remain unchanged.

The contributions of the four-quark condensates to the lhs of the Borel
transformed sum rules (\ref{57})--(\ref{59})  can be presented
in the same way, as for the symmetric case
\begin{eqnarray} &&
\omega^j=\omega^j_N\rho\ , \quad \omega^j_N=A^j_{4q}(\beta)f^j_{4q}\ ,
\nn \\
&&  f^q_{4q}=-8\pi^2 a , \quad f^p_{4q}=-8\pi^2\frac{s-m^2}{2m} a ,
\quad f^I_{4q}=-8\pi^2 m a.
\label{108}
\end{eqnarray}

\begin{center}
{\bf IV. SOLUTIONS OF THE SUM RULES EQUATIONS}
\end{center}

Now we present the solutions of the sum rules equations, focusing on
the functions $\Sigma_v(\rho,\beta)$ and $m^*(\rho,\beta)$. We shall
include the terms $\ell^j,u^j$ and $\omega^j$ (Eq.~(\ref{61})) in
the succession in lhs of Eqs.~(\ref{57})--(\ref{59}). Two lowest
order OPE contributions to the vector structures $\hat q$ and
$\hat p$ are presented in terms of the vector and gluon
condensates and of the nucleon structure functions. These
characteristics are either calculated in a model-independent
way, or determined in the experiments. The lowest order OPE
terms in the scalar channel are expressed in terms of
isotope--symmetric and isotope--asymmetric scalar condensates
$\kappa_N=\langle p|\bar uu+\bar dd|p\rangle$ and $\zeta_N=\langle
p|\bar uu-\bar dd|p\rangle$ --- Eqs.~(\ref{15}) and (\ref{16}). Here the
situation becomes somewhat more complicated.

The expectation value $\kappa_N$ is related to the $\pi N$ sigma-term
$\sigma_{\pi N}$ by Eq.~(\ref{17}). The value of $\sigma_{\pi N}$ can be
extracted from the data on low-energy $\pi N$ scattering. The
procedure consists in subtracting the high-order
chirality-violating terms $\sigma'$ from the experimental value
$\Sigma_{\pi N}$, i.e., $\sigma_{\pi N}=\Sigma_{\pi N}-\sigma'$. The
value $\sigma'\approx15\,$MeV was obtained in \cite{29} by
 dispersion relation technique. However, there are some
uncertainties in deducing the value of $\Sigma_{\pi N}$ from the
experimental data.  The canonical value $\Sigma_{\pi N} = (60\pm8)$MeV
\cite{12} is now challenged by the higher values $77\pm6\,$MeV
\cite{13}.  Assuming $m_u+m_d=11\,$eV \cite{11}, we find that
$\Sigma_{\pi N}=64\,$MeV corresponds to $\kappa_N=8$. Additional
uncertainties emerge, since the true value of the sum $m_u+m_d$ may be
somewhat larger.

There is no experimental data on the expectation value $\zeta_N$. If
the nucleon is treated as a system of the valence quarks and the
isospin symmetric sea of the quark--antiquark pairs, the expectation
value $\zeta_N$ is determined by the contribution of the valence
quarks. Thus, the reasonable values are $\zeta_N=1$ for the
nonrelativistic models and $\zeta_N<1$ in the relativistic case.
Until we include only the leading OPE terms $\ell^j$, we can
solve the sum rules equations for any values of $\kappa_N$ and
$\zeta_N$. However the four-quark condensates are obtained in framework
of a specific perturbative chiral quark model (PCQM). Within this model
$\sigma_{\pi N}=45\,$MeV \cite{20}, leading to $\kappa_N=8$. Thus, to
be self-consistent, we must use this value as the basic one in the
general equations, which include the contributions $\omega^j$. Note
also, that the values of $\Sigma_{\pi N}$ extracted from the
experimental data are correlated with the assumption on the strange
quark content $y_N=\frac{2\langle p|\bar ss|p\rangle}{\langle
p|\bar uu+\bar dd|p\rangle}$ \cite{13}. The values $\Sigma_{\pi
N}\approx 77\,$MeV correspond to $y_N\approx0.35$, with a large
part of the nucleon mass being due to the strange quarks. The
smaller values of $\Sigma_{\pi N}$ require much smaller values
of $y_N$.  In PCQM one finds $y_N=0.08$ \cite{20}, in agreement with the
smaller values of $\Sigma_{\pi N}$. The PCQM value
$\zeta_N=0.54$ can be obtained by using the results of
\cite{19}.

We shall present most of the numerical results for the values
\begin{equation}
\label{109}
\kappa_N\ =\ 8\ , \quad \zeta_N\ =\ 0.54\ .
\end{equation}
Anyway, in the linearized version of the sum rules, the nucleon
characteristics will be presented as the explicit function of the
condensates, e.g., of the parameters $\kappa_N$ and $\zeta_N$.

\begin{center}
{\bf A. Solution of the general equations}
\end{center}

Here we present the solutions of the general equations
(\ref{57})--(\ref{59}), which are identical to Eqs.
(\ref{52})--(\ref{54}). Recall that we approximate the in-medium
condensates by the functions, which are linear both in $\rho$ and
$\beta$.  However the solutions $\Sigma_v(\rho,\beta)$ and
$m^*(\rho,\beta)$ are not linear. One can demonstrate this by
presenting Eqs.~(\ref{53}) and (\ref{54}) as
\begin{equation}
\label{110}
\Sigma_v\ =\ -\frac{L^p_m}{L^q_m}\ , \quad m^*\ =\
\frac{L^I_m}{L^q_m}\ ,
\end{equation}
with the density and $\beta$ dependence of $L^q_m$ leading to nonlinear
behavior of $\Sigma_v$ and $m^*$ (even if we assume $W^2_m=W^2_0$). The
nonlinear dependence of rhs of Eq.~(\ref{110}) on $W^2_m$ also cause
the nonlinear contributions to $\Sigma_v$ and $m^*$. However, they are
numerically less important.

Now we include the terms $\ell^j,u^j$ and $\omega^j$ in
succession in lhs of Eqs.~(\ref{57})--(\ref{59}).

\begin{center}
{\bf {\it 1. The role of the lowest order local condensates}}
\end{center}

As we have seen, these are the contributions, which contain the vector
condensate $v(\rho)$, gluon condensate $g(\rho)$ and the scalar
condensates $\kappa_m$ and $\zeta_m$ --- Eqs.~(\ref{11}),
(\ref{13}), (\ref{14}), (\ref{18}). Account of these terms
corresponds to the one-meson exchanges between the nucleon under
consideration and the nucleon of the matter, with the point-like
structures of the meson--nucleon vertices, Fig.~2(a). The solution
can be obtained by using Eqs.  (\ref{68})--(\ref{70}) for the
functions $\ell_j$. As we have seen, there is a simple solution,
expressed by Eqs.~(\ref{73})--(\ref{75}).  The procedure of
minimization of the difference between the lhs and rhs of Eqs.
(\ref{57})--(\ref{59}) indeed prefers the values
$W_m(\rho,\beta)\approx W_m(\rho,0)$. Thus Eqs.~(\ref{73}) and
(\ref{74}) appear to be true with good accuracy. Hence,
\begin{eqnarray}
&&
\Sigma^{(p)}_v(\rho,\beta)=\Sigma_v(\rho,0)\left(1-\frac\beta4\right)\
, \nn\\ &&
\Sigma^{(n)}_v(\rho,\beta)=\Sigma_v(\rho,0)\left(1+\frac\beta4\right)\
, \nn\\ \label{111}
&&
m^{*(p)}(\rho,\beta,\kappa_N,\zeta_N)=
m^*(\rho,0,\kappa_N+\beta\zeta_N,0)\ ,
\nn\\
&&
m^{*(n)}(\rho,\beta,\kappa_N,\zeta_N)=
m^*(\rho,0,\kappa_N-\beta\zeta_N,0)\ .
\end{eqnarray}
Thus, in the matter with the excess of the neutrons $(\beta>0)$, we
obtain $\Sigma^{(n)}_v>\Sigma^{(p)}_v$, and $m^{*(n)}>m^{*(p)}$. For
example, using the value $\Sigma_v(\rho,0)$ obtained by sum rules
approach in \cite{16} ($\Sigma_v(\rho,0)=335$~MeV), we find
$\Sigma^{(n)}_v-\Sigma^{(p)}_v=170\,$MeV, $m^{*(n)}-m^{*(p)}=50\,$MeV
for the neutron matter $(\beta=1)$ at $\rho=\rho_0$.

The minimization procedure choses $W^2_m(\beta=-1)=2.50\,\rm
GeV^2$, $W^2_m(\beta=0)=2.30\,\rm GeV^2$,
$W^2_m(\beta=1)=2.05\,\rm GeV^2$,
$\Sigma^{(n)}_v-\Sigma^{(p)}_v=170\,$MeV and $m^{*(n)}-
m^{*(p)}=50\,$MeV. Thus, Eqs.~(\ref{111}) work well indeed. However,
this approximation is not sufficient for the description of the
potential energy $U(\rho,\beta)$ (\ref{24}), providing $U>0$
for both symmetric and asymmetric cases.

\begin{center}
{\bf {\it 2. The role of the four-quark condensates}}
\end{center}

Now we include the four-quark condensate, i.e., we use Eq.~(\ref{60})
for $L^j = \ell^j + \omega^j$, with $\omega^j$ described by Eqs.
(\ref{108}).  Inclusion of these terms mimics several contributions on
the rhs of the sum rules (\ref{57})--(\ref{58}). In the condensates
$\bar u\Gamma^Xu\bar u\Gamma^Xu$, presented by Eq.~(\ref{101}) the
first term, which has a nonvanishing value only if $\Gamma^X=I$,
generates a contribution to the $\hat q$ structure due to the
anomalous Lorentz structure of the interaction between the
scalar field and the nucleon, caused by the chiral-odd vacuum
condensate $\langle 0|\bar qq|0\rangle$. In similar way the first
term on the rhs of Eq.~ (\ref{102}) describes the contribution
of the vector meson exchange to the scalar structure of the
nucleon propagator.  The anomalous Lorentz structures emerge if
the nucleon-meson vertices are treated beyond the lowest order.
These contributions are illustrated by Fig.~2(b). The second terms
on the rhs of Eqs.  (\ref{101}) and (\ref{102}) and on the rhs
of Eq.~ (\ref{103}) describe  exchanges by the four-quark
strongly correlated system (see Fig.~2(c)).
The condensates presented by Eq.~(\ref{103}) have the same
values for the proton and neutron. Thus their contributions do
not depend on $\beta$.

The differences between the neutron and proton characteristics
in the neutron matter at $\rho=\rho_0$ become
$\Sigma^{(n)}_v-\Sigma^{(p)}_v=140\,$MeV, $m^{*(n)}-m^{*(p)}=-110\,$MeV.
In the simplified model for the pion with the current masses of the
constituent quarks, where $A^q_{4q}(\beta)$ is given by Eq.
(\ref{107}), we obtain $\Sigma^{(n)}_v-\Sigma^{(p)}_v=145$~MeV, while
$m^{*(n)}-m^{(p)}=-110$~MeV at these values of $\rho$ and $\beta$.

\begin{center}
{\bf {\it 3. The final results}}
\end{center}

These contributions come from the account of the $x$-dependence of the
expectation values of the vector operators $\langle M|\bar
q^i(0)\gamma_\mu q^i(x)|M\rangle$ with $q^i(x)$ defined by Eq.
(\ref{63}). As we have seen, the nonlocality of the scalar condensates
is not important in our case. The nonlocality is included by putting
$L^j=\ell^j+\omega^j+u^j$ with $u^j$ defined by Eq.~(\ref{85}). We use
the structure functions, obtained in \cite{30} for the calculation of
the terms $u^q$ and $u^p$.

Account of the nonlocality of the vector condensate corresponds to
inclusion of the formfactor of the vertex of the interaction between
the vector meson and the nucleon of the matter - Fig.~2(d). Recall
that similar contribution for the effective scalar meson
exchanges vanishes in our approximation - Sec. III.B.

We find the dependence of the nucleon vector self-energy
$\Sigma_v$ and of the effective mass $m^*$ on density of the
matter and on the asymmetry parameter $\beta$ and show the
results in Figs.~3 and 4. For example, the differences between
the neutron and proton characteristics in the neutron matter at
$\rho = \rho_0$ are
$\Sigma^{(n)}_v - \Sigma^{(p)}_v=110$~MeV,
$m^{*(n)} - m^{*(p)}=-70$~MeV.
In the simplified model for the
pion with the current quark masses, where
$A^q_{4q}(\beta)$ is given by Eq.~(\ref{107}) we obtain
$\Sigma^{(n)}_v - \Sigma^{(p)}_v=115$~MeV,
$m^{*(n)} - m^{*(p)}=-65$~MeV at these values of $\rho$ and
$\beta$.
The minor change is due to the small change in the
$\beta$ dependence of the condensate $A^q_{4q}$.

Due to the nonlinear character of Eqs.~(\ref{57})-(\ref{59}) the
self-energies $\Sigma_v(\rho,\beta)$ and $\Sigma^*_s(\rho,\beta)
= m^*(\rho,\beta)-m$ could have been nonlinear in both $\rho$
and $\beta$.  The nonlinear behavior  of these characteristics
with $\rho$ manifests itself explicitly. However, the dependence
on $\beta$ appears to be linear in framework of the accuracy of
our computations (see the next Subsection). Thus both $\Sigma_v$
and $\Sigma^*_s$ can be approximated by linear functions of
$\beta$ \begin{eqnarray}
&& \Sigma_v(\rho,\beta)=
\frac\rho{\rho_0}\bigg(V_1(\rho)+\beta\tau_zV_2
(\rho)\bigg),
\nn\\
&&
\Sigma^*_s(\rho,\beta)
=\frac\rho{\rho_0}\bigg(S_1(\rho)+\beta\tau_z S_2(\rho)\bigg)
\label{112}
\end{eqnarray}
with $\tau_z=1$ for the proton, $\tau_z=-1$ for the neutron. The
functions $V_{1,2}(\rho)$ and $S_{1,2}(\rho)$  are shown in Fig.~5.
They can be approximated by polynomials of the second order -- see
Appendix D.

The single-particle potential energy is expressed by
Eq.~(\ref{24}). At $\rho=\rho_0$ the neutron-proton difference of
the potential energy caused by the isovector interaction is
$\Delta U_{np} \approx 38\beta$~MeV at small $\beta$. In Fig.~ 6
we show the dependence $U(\rho)$ for several values of $\beta$
for both neutrons and protons. Recall that the potential energy
is determined with the lower accuracy then the self-energies.

The nucleon residue $\lambda^2_m$ and the spectrum threshold $W^2_m$
exhibit very weak dependence on $\beta$. Thus we can assume
\begin{equation}
\label{116}
\lambda^2_m(\rho,\beta)\approx\lambda^2_m(\rho,0), \quad
W^2_m(\rho,\beta)=W^2_m(\rho,0)\ .
\end{equation}

\begin{center}
{\bf B. Explicit expression for the nucleon parameters in
terms of QCD condensates}
\end{center}

We can present an approximate solution of Eqs.~(\ref{57})--(\ref{59}),
in which the nucleon self-energies are expressed in terms of the QCD
condensates explicitly. We see from Eq.~(\ref{116}), that
$W^2_m(\rho,\beta)\approx W^2_m(\rho,0)$,
while $W^2_m(\rho,0)$ is close to it's vacuum value $W^2_0$
Eq.~(\ref{46})\cite{16}.  Thus, we can put $W^2_m=W^2_0$ in the rhs of
Eqs.  (\ref{52})-(\ref{54}) and (\ref{110}). This enables us to
present the proton characteristics $\Sigma_v$ and $m^*$ as the
explicit functions of the quark condensates and of the Borel
mass $M^2$:
\begin{eqnarray}
\Sigma_v&=&-
\frac{T^p_v\left(v_N+\frac34\beta v_N^{(-)}
\right)+T^p_{u1}+\beta T^p_{u2}
+T^p_\omega A^p_{4q}(\beta)}{1+F^q(\rho,\beta)}\frac\rho{\rho_0}\ ,
\nonumber\\
&&
\label{117}\\
m^*&=& \left[m+\left(T^I_\kappa(\kappa_N+\beta\zeta_N)
+T^I_\omega A^I_{4q}(\beta)\right)\frac\rho{\rho_0}\right]
\frac{1}{1+F^q(\rho,\beta)}
\label{118}
\end{eqnarray}
with
\begin{equation}
F^q(\rho,\beta) =
\bigg[T^q_v v_N+mT^q_g g_N+T^q_{u1}
 +\ \beta T^q_{u2}+T^q_\omega A^q_{4q}(\beta)\bigg]
\frac\rho{\rho_0}\ .
\end{equation}
We denote $T^i_\omega=T^i_\omega(M^2)$ and
$T^i_j=T^i_j(M^2,W^2_0)$ for other $j$
and introduce for $i=q,p,I$
\begin{eqnarray}
&& T^i_k(M^2,W^2_0)=\rho_0f^i_k(M^2,W^2_0)
\frac{e^{m^2/M^2}}{\lambda^2_0}\ , \quad (k=v,g,\kappa)\ ,
\nonumber\\
&& T^i_{ur}(M^2,W^2_0)=\rho_0u^{i}_{N,r}(M^2,W^2_0)
\frac{e^{m^2/M^2}}{\lambda^2_0}\ , \quad (r=1,2)\ ,
\nonumber\\
&& T^i_\omega(M^2)\ =\ \rho_0 f^i_{4q}\
\frac{e^{m^2/M^2}}{\lambda^2_0}
\label{120}
\end{eqnarray}
with the functions $f^q_k$ and $f^q_{4q}$ defined by Eqs.~(\ref{72}) and
(\ref{108}). It is instructive to present the density $\rho$ in
units of the observable saturation density of the symmetric
matter $\rho_0=0.17\,\rm fm^{-3}$.

Note that the functions $T^i_j(M^2)$ defined by Eq.~(\ref{120})
$(j=v,g,\kappa,u1,u2; i=q,p,I)$ depend on $M^2$ rather weakly.
Thus, approximating
\begin{equation} \label{126}
T^i_j(M^2)\ =\ C^i_j\ ,
\end{equation}
we can replace
the functions $T^i_j(M^2)$
in the lhs of Eqs.~(\ref{120})
by the constant coefficients $C^i_j$.
Numerically most important functions $T^p_v(M^2)$ and $T^I_\kappa(M^2)$
can be approximated by the constant values with the errors of about
4\% and 7\%. The largest errors of about 25\% emerge in the averaging
of the  functions $T^i_\omega$. This solves the
problem of expressing the in-medium change of nucleon parameters
through the values of the condensates. For the proton
\begin{equation}\label{127a}
\Sigma_v = -\left(C^p_v v_N+\beta C^p_{v^{(-)}}v^{(-)}_N +
mC^p_{u1} + \beta m C^p_{u2}+mC^p_\omega
A^p_{4q}(\beta)\right)\frac\rho{\rho_0}
\frac1{(1-{\cal F}_q)}\ ,
\end{equation}
\begin{equation}\label{127}
m^* = \left[m +
\left(C^I_\kappa\kappa_N+\beta C^I_\zeta\zeta_N
 +C^I_\omega A^I_{4q}(\beta)\right)\frac\rho{\rho_0}
\right]\frac1{(1-{\cal F}_q)}\ ,
\end{equation}
with
\begin{eqnarray}\label{128}
{\cal F}_q = -\left(
C^q_vv_N+mC^q_gg_N+mC^q_{u1}+ \beta mC^q_{u2}
+mC^q_\omega A^q_{4q}(\beta)\right)\frac\rho{\rho_0}.
\end{eqnarray}
The coefficients in the rhs of Eqs.~(\ref{127a})--(\ref{128}) are
\begin{eqnarray}
&&
C^q_v=-0.062, \quad C^q_g=0.011\mbox{ GeV}^{-1}, \quad
C^q_\omega=-0.070,
\nn \\
&&
C^q_{u1}=-0.074, \quad C^q_{u2}= 0.008 ,
\nn \\
&&
C^I_\kappa=-0.042\mbox{ GeV, }\ C^I_\zeta=-0.042\mbox{ GeV, }
C^I_\omega=-0.063\mbox{ GeV },
\nn \\
&&
C^p_v=-0.090\mbox{ GeV},\  C^p_{v^{(-)}}=-0.068\mbox{ GeV, }\
C^p_\omega=-0.095\ ,
\nn \\
&&
C^p_{u1}=0.094\ ,\quad C^p_{u2}=-0.020 .
\label{129}
\end{eqnarray}

Note that the dependence of ${\cal F}_q$ on $\beta$ is very weak.
(Recall that the lowest order OPE terms in the $\hat q$
structure of polarization operator do not depend on $\beta$.)
Thus, in the lhs of Eqs.~(\ref{127a}), (\ref{127})
only the dependence of the
numerators on $\beta$ is important. This explains the linear
dependence of the self-energies $\Sigma_v$ and $\Sigma^*_s$ on
$\beta$.

The values of $\Sigma_v$, $m^*$ and ${\cal F}_q$ for the neutron are
described by Eqs.  (\ref{127a})--(\ref{128}) with $\beta$ changed
to $-\beta$.  Equations (\ref{127a}), (\ref{127})  enable to
obtain $\Sigma_v$ and $m^*$ during the successive inclusion of
condensates of the higher dimensions. If only the leading OPE
terms are included the values provided by Eqs.~(\ref{127a}),
(\ref{127}) actually coincide with the solutions of Eqs.
(\ref{57})--(\ref{59}). If all the contributions are included
Eqs.~(\ref{127a}), (\ref{127}) reproduce the values of $\Sigma_v$
and $m^*$ with the accuracy of 15$\%$ and 10$\%$ correspondingly
for the symmetric matter.  The precision of Eqs.  (\ref{127a})
and (\ref{127}) changes with $\beta$.  For the neutron matter
these equations   provide the values $\Sigma^{(n)}_v -
\Sigma^{(p)}_v=120$~MeV and $m^{*(n)} - m^{*(p)}=-60$~MeV,
comparing to the values of 110~MeV and -70~MeV, obtained in the
previous subsection from the general solution of Eqs.
(\ref{57})--(\ref{59}).

\begin{center}
{\bf V. DISCUSSION}
\end{center}

Now we compare our results to those, obtained by nuclear physics
methods. The lowest order OPE terms in the lhs of the sum rules
describe mainly the exchanges by the localized $\bar qq$ pairs.
This corresponds to the vector and (effective) scalar meson
exchanges between  the nucleon and the  nucleons of matter.
These exchanges have the point-like vertices and the standard
Lorentz structures.

Inclusion of the higher order OPE terms corresponds to a more
complicated picture of the meson exchanges between the
nucleons. Turning to the four-quark condensates, we can separate
the two types of terms.  In the "factorized" contributions one
of the $\bar qq$ operators is averaged over vacuum. In the
"internal" terms both $\bar qq$ pairs act inside the nucleons ,
Eqs.~(\ref{89}), (\ref{90}), (\ref{101})-(\ref{103}). The first
(factorized) term in the rhs of Eq.~(\ref{101}) describes the
contribution to the $q_\mu\gamma^\mu$structures of the
polarization operator, which contains the scalar expectation
value $\langle N|\bar uu|N\rangle$. In similar way the first
(factorized) term in the rhs of Eq.~(\ref{102}) contributes to
the scalar structure of the polarization operator, being
proportional to the vector expectation value
 $\langle N|\bar u\gamma_\mu u|N\rangle$. These terms correspond
to the anomalous Lorentz structures of the nucleon-meson
vertices. As to the "internal" terms, i.e.,  the last terms on the
rhs of Eqs.~(\ref{89}), (\ref{90}), they can be interpreted as
the exchanges by two-meson systems with their local interactions
with the nucleon or as the exchanges by four-quark mesons (if
there are any).

Inclusion of the nonlocal vector condensates $\bar
q(0)\gamma_\mu q(x)$ means that the vertices of the interactions
between the nucleons of the matter and the vector mesons do
not have a point-like structure, requiring rather description
by the formfactors. The nonlocality of the scalar condensate does
not influence the results in our approach.

A usual subject of calculation is the difference between the
characteristics of neutron and proton. If only the lowest OPE
terms are included, the vector self-energies are determined by
the vector condensates. The neutron-proton difference
$\Sigma_v^{(n)}- \Sigma_v^{(p)}$, usually attributed to the
$\rho$ meson exchange is 170~MeV at $\rho=\rho_0$ and $\beta=1$.
Inclusion of the four-quark condensates and of the nonlocalities
subtract 30~MeV and 28~MeV from this value. The lowest order OPE
terms provide the difference of the effective masses
$m^{*(n)} - m^{*(p)}$ = 50~MeV at the same values of $\rho$ and
$\beta$.
Inclusion of the four-quark condensates and of the nonlocalities
adds (-160~MeV) and 40~MeV. This leads to
$\Sigma_v^{(n)}- \Sigma_v^{(p)} = 110$~MeV and
$m^{*(n)} - m^{*(p)}$ = -70~MeV in neutron matter at
$\rho=\rho_0$

The structure of the equations for $\Sigma_v$ amd $m^*$
(\ref{127a}), (\ref{127}) is similar to that (\ref{23}) employed in nuclear
physics.
Recall that in the Hartree approximation dependence
 of $\Sigma_v$ on the density deviates from the linear
law due to $\Sigma_q$ (\ref{23}).
(The same refers  to $\Sigma^*_s$ if the nucleon
Fermi motion is neglected). In our approach the
nonlinear behavior of the self energies is due
to nonzero values of ${\cal F}_q$.

Now we compare the numerical results.
Considering the papers, containing relativistic calculations,
we can compare the vector and scalar self-energies $\Sigma_v$ and
$\Sigma^*_s=m^*-m$. In the case of the works, carried out in the
nonrelativistic approximation, we can compare the nucleon potential
energies $U^{n,p}$. We analyze also the contribution to the parameter,
conventionally denoted as $a_4$ \cite{FW}, which is defined as
\begin{equation}
\label{130}
\varepsilon(\rho_0,\beta)\ =\ \varepsilon(\rho_0,0)+\beta^2a_4
+O(\beta^4)\ ,
\end{equation}
being thus the lowest order term of $\beta^2$ expansion of the
averaged binding energy $\varepsilon$ per nucleon at saturation value
of density.

Of course, we cannot expect very good agreement, since our calculations
are carried out in the gas approximation. The future and more
sophisticated calculations should include the scalar and four-quark
condensates beyond the gas approximation. This would correspond to the
account of the renormalization of the nucleon interactions with the
matter by the particle-hole excitations in rhs of the sum rules.
Another reason is that the results should be corrected for the effects
of antisymmetrization of the total final state wave function
("exclusion effect") \cite{31}.

The general feature of the relativistic calculations is that they
provide the positive value of the difference
$\Sigma^{(n)}_v-\Sigma^{(p)}_v>0$ in the matter with the neutron
excess. Also the proton effective masses are above the neutron
ones in this case, i.e., $m^{*(n)}-m^{*(p)}<0$.  Our calculations
show the same tendency. As to the quantitative results, our
values of $\Sigma^{(n)}_v-\Sigma^{(p)}_v$ and
$m^{*(n)}-m^{*(p)}$ appear to be about twice smaller than those,
obtained in \cite{32}. Our value of the effective mass splitting is
also about two times smaller than the result of \cite{33} but is only
30\% smaller than that of \cite{34}. However, we find somewhat smaller
discrepancy with the relativistic Brueckner--Hartree--Fock (RBHF)
calculations, presented in \cite{35}. They found
$\Sigma^{(n)}_v-\Sigma^{(p)}_v\approx30\,$MeV and
$m^{*(n)}-m^{*(p)}\approx-25\,$MeV at $\beta=0.2$ (Fig.~3 of \cite{35}),
while our results are $\Sigma^{(n)}_v-\Sigma^{(p)}_v\approx20\,$MeV,
$m^{*(n)}-m^{*(p)}\approx-15\,$MeV. Another RBHF analysis \cite{36}
provided the results, which are very close to our ones. One can extract
the values $\Sigma^{(n)}_v-\Sigma^{(p)}_v\approx80\,$MeV,
and $m^*(n)-m^*(p)\approx-50\,$MeV
 at $\beta=0.75$ from Figs.~10 and 11 of \cite{36}.
 Our values are 80~MeV
and 55~MeV correspondingly. Note, however, that the split of the
effective masses, obtained in \cite{36} is due to the exchange
effects only.

The nonrelativistic calculations, carried out in various
approaches \cite{37,38} provide $U^{(n)}-U^{(p)}\approx60\,$MeV
at $\rho=\rho_0$ and $\beta=1$.  This is consistent with the
earlier calculations \cite{39}. Our value is 40~MeV at
$\rho=\rho_0$ and $\beta=1$.

Another important parameter is the symmetry energy --- Eq.
(\ref{130}). Note that we calculate the quantity
\begin{equation}
\label{131}
\Delta\,\varepsilon\ =\
\frac1{2\rho}\left(U^{(n)}\rho_n+U^{(p)}\rho_p\right),
\end{equation}
which is the true contribution  to the energy per
nucleon. Using Eq.~(\ref{112}) we can express $\Delta\varepsilon
= \frac12 (V_1 + S_1 - \beta^2(V_2 + S_2))$. The value
\begin{equation} \label{132}
\Delta\,\varepsilon(\rho_0,\beta)-\Delta\,\varepsilon(\rho_0,0)\
=\ \beta^2\Delta a_4+O(\beta^4)\ ,
\end{equation}
thus being the true contribution of the isovector forces. Our value
 is $\Delta a_4\approx 10$~MeV.
This is close to the one, obtained in \cite{31}. The
 $\beta^2$ law is true up
to $\beta^2=1$ with 10\% accuracy in agreement with
\cite{35,37,38}.  To find the total contribution of the
potential energy one must include the term caused by the exclusion
effect, mentioned above.  This adds 9~MeV to $a_4$
\cite{31}. Including also the contribution of the kinetic
energy, we obtain $a_4=29\,$MeV.  The various calculations of
 this parameter provide the values around 30~MeV
\cite{32}--\cite{40}. Thus, our result agrees with those,
obtained by nuclear physics methods.

\begin{center}
{\bf V. SUMMARY}
\end{center}

We  expressed the vector and scalar self-energies of a nucleon in
asymmetric nuclear matter as function of density $\rho$ and of
the asymmetry parameter $\beta$. We presented the nucleon
characteristics in terms of the in-medium expectation values of
QCD operators. The main ingredients are the nonlocal vector
condensates $\langle M|\bar u(x)\gamma_0u(0)\pm\bar
d(x)\gamma_0d(0)|M\rangle$, the scalar condensates $\langle
M|\bar u(0)u(0)\pm\bar d(0)d(0)|M\rangle$ and the four-quark
condensates. The local vector condensates are calculated easily.
The nonlocality of the vector condensates is expressed in terms
of the nucleon structure functions. The scalar condensate
$\kappa(\rho)=\langle M|\bar uu+\bar dd|M\rangle$ is presented
in terms of the observable sigma term. The scalar condensate
$\zeta(\rho,\beta)=\langle M|\bar uu-\bar dd|M\rangle$ and the
four-quark condensates are calculated in framework of perturbative
chiral quark model (PCQM).

Although we treat the condensates in the gas approximation, the nucleon
characteristics are not linear in density.
The corresponding
equations (\ref{127a}), (\ref{127}) are analogous to the
equations of nuclear physics beyond the mean field
approximation. Also Eqs.~(\ref{127a}), (\ref{127}) provide
explicit expression of the nucleon characteristics in terms of
the QCD condensates.

The successive inclusion of the OPE terms in the lhs of the sum
rules finds direct analogs in the meson-exchange description of
the interactions of the nucleon in nuclear matter. The lowest
order OPE terms correspond to the exchanges by the vector and
(effective) scalar mesons with the point-like vertices of the
interactions. The higher order terms correspond to the nonlocal
structure of nucleon-meson vertices, including the anomalous
Lorentz structures, and to the exchanges by the strongly
correlated four-quark systems. A possible interpretation of the
latter contributions is a local two-meson exchange (or the
exchanges by the four-quark mesons, if there are any \cite{41}),
Figs.~1, 2.

We obtained the functions $\Sigma_v(\rho,\beta)$ and
$m^*(\rho,\beta)$ for all values of $\beta$.
  We calculated also the single-particle potential energy
$U(\rho,\beta)$.  The results are presented in Figs.~3-6.

Note, that we did not need phenomenological parameters of the
nucleon--meson interactions. We used the condensates which have
been either calculated, or expressed in terms of the
observables.  For the four-quark condensates we used, however,
the already known input parameters of PCQM. While including the
condensates of the higher dimension in succession, we found the
direct analogs of the meson--nucleon exchange mechanisms of
nuclear physics.

Our results for the nucleon self-energies are in reasonable agreement
with the results of nuclear physics. The value of the symmetry energy
is close to the one obtained by the nuclear physics methods.

The work was supported by the
grants RFBR - 03-02-17724 and RSGSS - 1124.2003.2.

\begin{center}
{\bf APPENDIX A}
\end{center}

Here we discuss the isotopic structure of the nucleon current
(\ref{28}).

The current (\ref{28})  does not have a definite isospin.
This is not important for the calculations in a medium with the isospin
I=0, e.g., in vacuum or in symmetric nuclear matter. In
these cases the expectation values of $\bar uu$ and $\bar dd$
operators have equal values (the same is true for more
complicated condensates). This reasoning does not work for
asymmetric matter.

Therefore it is reasonable to start with
the pure I=1/2 current
$$
j(y)
=\epsilon_{abc}\cdot\left[\sqrt{\frac 23} u^a(y)C\gamma_\mu
u^b(y)\gamma_5\gamma^\mu d^c(y)\right.
$$
$$
\left.-\frac 1{\sqrt 6} u^a(y)C\gamma_\mu d^b(y)\gamma_5\gamma^\mu
u^c(y) -\frac 1{\sqrt 6}  d^a(y)C\gamma_\mu
u^b(y)\gamma_5\gamma^\mu u^c(y)\right]
\eqno{(A1)}
$$
However it turns out that even in asymmetric matter an explicit
calculation of the polarization operator by using the current
(A1) provides the same result
as that with the original Ioffe current (\ref{28}).

This is due to the Fermi statistics of the quarks.
In the OPE the current (\ref{28})
annihilates a local three-quark state.
The coordinate function of the three quarks is symmetric. The
color structure is antisymmetric
as well as the spin structure of the spin three quark system
(clearly the current (\ref{28}) has spin 1/2). Thus we need the
antisymmetric isospin wave function. Due to the Fermi statistic
of the fermions any symmetric isospin component will
vanish after the account for all possible permutations of the quarks.
This means that only the isospin 1/2 component actually survives
in the OPE expansion of the polarization operator (\ref{27})
corresponding to the of current (\ref{28}).  In other words
there is a relation between the contributions of the first and
two last terms of the rhs of Eq.~(A1). This provides a
possibility to use (28) for the calculations in
isospin asymmetric matter.

\begin{center}
{\bf APPENDIX B}
\end{center}

The integrals in the rhs of  Eq.~(\ref{84}) can be evaluated by
using the formula
$$
\int\frac{d^4x}{x^8}x_\mu x_\nu e^{i(q'x)}
= \frac 16\left[g_{\mu\nu} + \frac{2g'_\mu
q'_\nu}{q'^2}\right]\int\frac{d^4x}{x^6}e^{i(q'x)} \eqno{(B1)}
$$
with $q'=q-p\alpha$. The last factor is
$$
\int\frac{d^4x}{x^6}e^{i(q'x)} = -\frac{i\pi^2}8 q'^2\ln(-q'^2)
+ . . .
\eqno{(B2)}
$$
Here the dots denote the terms which will be killed by the Borel
transform.

Thus the lhs of the sum rules contains the contribution of the
form
$$
X = \int d\alpha\ln(Q^2+A^2(\alpha)) f(\alpha)
$$
with $Q^2=-q^2$,
$A(\alpha) = \alpha\frac{(s-m^2) - m^2\alpha}{1+\alpha}$. We can
present
$$
X =
\ln Q^2\int\limits^1_0 d\alpha f(\alpha)\
+\int\limits^1_0 d\alpha\ln\frac{Q^2+A^2(\alpha)}{Q^2}\,f(\alpha) .
  \eqno{(B3)}
$$
The first term on the rhs contains the standard logarithmic factor
with the cut, running to infinity. It is described by the
"pole + continuum" model in a standard way. The second term
contains a finite cut.  It describes the singularities in the
$u$-channel, caused by the nonlocal structure of the vector
condensate, corresponding to the exchange terms on the rhs of
the sum rules. We neglect such contributions, thus coming to the
Hartree description of the nucleon in nuclear matter.

\begin{center}
{\bf APPENDIX C}
\end{center}

Here we calculate the contribution of the four-quark condensates
to $I$ structure of the polarization operator $\Pi_m(q)$. The
contribution is provided by the scalar-vector condensate
$r^{SV}_N$ defined by Eq.~(\ref{90}) with $X=S, Y=V$. The first
two terms determine the "factorized" contribution, which can be
written as
$$
\Pi^I_f = -\frac 23\int\frac{d^4}{\pi^2x^4}(x,\theta^u(x)) e^{i(qx)}
\langle 0|\bar dd|0\rangle ,
\eqno{(C1)}
$$
with the vector $\theta^u_\mu(x)$ defined by Eq.~(\ref{64}).
Neglecting for a while the nonlocal contributions, we put
$\theta^u(x) = \theta^u(0)$. In this approximation
$$
\Pi^I_f  = \Pi^I_{f,loc} = -\frac 43\frac{(pq)}{q^2m}
\langle 0|\bar dd|0\rangle (2\rho_p+\rho_n),
\eqno{(C2)}
$$
thus corresponding to Eq.~(\ref{100}) with
$a^{SV}_p = \frac 43\langle 0|\bar dd|0\rangle$,
$a^{SV}_n = \frac 23\langle 0|\bar dd|0\rangle$. The "internal"
contributions, provided by the second terms in the rhs of
Eqs.~(\ref{90}) and (\ref{102}) are also proportional to $(pq)$,
with  the coefficients $(a^{SV}_i)_{int}$ determined in
\cite{16}. The factorized local terms and the internal terms
provide the rhs of the last equality of Eq.~(\ref{100}).

Taking into account the $x$-dependence of the vector
$\theta^u_\mu(x)$ in  Eq.~(C1) we obtain a more
accurate expression for the "factorized" contribution, which
contains the second moments of the structure functions and the
terms of the $x^2$ expansion of the integrand
$$
\Pi^I_d = \frac{\langle 0|\bar dd|0\rangle }{q^2}\frac m3\left[
-4\frac{(pq)}{m^2}
(\langle \eta_u\rangle\rho_p +
\langle \eta_d\rangle\rho_n) +
2(\langle \alpha\eta_u\rangle\rho_p +
\langle \alpha\eta_d\rangle\rho_n) \right.
\nn
$$
$$\left.
+(\langle \xi_u\rangle\rho_p +
\langle \xi_d\rangle\rho_n) +
\frac 12\frac {\mu_0^2}{m^2}\frac1{q^2}
(\langle \eta_u\rangle\rho_n +
\langle \eta_d\rangle\rho_p)\right] .
\eqno{(C3)}
$$
The notations are explained in Eqs.~(\ref{79}) and (\ref{81}),
$\mu_0^2 =0.8$~GeV$^2$ (\ref{41}), (\ref{42}).

\begin{center}
{\bf APPENDIX D}
\end{center}

The functions $V_{1,2}(\rho)$ and $S_{1,2}(\rho)$ introduced
 by Eq.~(\ref{112}) can
be approximated by the polynomials of the second order:
$$
P_2(x) = b_0 + b_1 x + b_2 x^2
\eqno{(D1)}
$$
with $x=\rho/\rho_0$. The values of the coefficients
$b_i$ are presented in Table 1.
This leads to  parametrization of the proton potential
energy
$$
U(x)=[x(-0.15 + 0.17x -0.09x^2) +
\beta x (-0.04x +0.02x^2)] \mbox{GeV} .
\eqno{(D2)}
$$

{\bf Table }
The values of the coefficients $b_i$ (GeV) of
the polynomials $P_2$ (D1) ,
which approximate the nucleon self-energies.

\begin{tabular}{|r|r|r|r|}
\hline
&
$b_0$
&
$b_1$
&
$b_2$
\\
\hline
$V_1$
&
$ 0.102$
&
$ 0.015$
&
$ 0.026$
\\
\hline
$V_2$
&
$-0.036$
&
$-0.008$
&
$-0.011$
\\
\hline
$S_1$
&
$-0.254$
&
$ 0.150$
&
$-0.114$
\\
\hline
$S_2$
&
$ 0.034$
&
$-0.035$
&
$ 0.034$
\\
\hline
\end{tabular}

\newpage

\section{Figure captions}
\noindent
Fig.1.
Successive inclusion of the OPE terms.
The helix lines denote the current (\ref{28}).
Solid lines denote the quarks.
Thick lines show the matter.
(a)~The local vector and
scalar condensates are included (contribution of the gluon
condensate is not shown).
(b)~The factorized  part of the
four-quark condensates is included. One of the $\bar qq$
operators acts on vacuum, while another $\bar qq$ operator acts
inside the nucleon.  Small circles stand for averaging over vacuum.
(c)~The internal parts of the four-quark
operators are included. All the quark operators act inside the
nucleon.
(d)~The nonlocality of the vector condensate is included.
The dark blobs denote the nonlocal condensates.

\noindent
Fig.2.
The analogs of the OPE terms in the meson-exchange interaction
of the nucleon with matter. Figures 2(a-d) correspond to the
Figures 1(a-d).
 Solid line denotes the nucleon.
Thick lines show the matter.
Wavy and dashed lines stand for vector and scalar mesons.
(a)~Exchanges by vector and scalar ("effective")
mesons with the pion vertices.
(b)~Exchanges  by these mesons
with the anomalous Lorentz structures of the vertices
(the dark squares).
(c)~Exchanges by the $\bar qq$ pairs with strong correlations
between them, presumably the local interactions with two mesons.
 Double line  denotes the two-meson systems.
(d)~Inclusion of the nonlocal structure of the vertices of
interaction between the vector mesons and the nucleons of the
matter. The dark blobs denote the formfactors of the vertices.

\noindent
Fig.3.
The density dependence of the vector self-energy
$\Sigma_v$ (solution of Eqs.~(\ref{57})-(\ref{59})). The solid
line shows the results for symmetric matter ($\beta=0$). The
dashed and dotted lines show the proton and neutron
self-energies $\Sigma_v^{(p)}$ and $\Sigma_v^{(n)}$
in the neutron matter ($\beta=1$).

\noindent
Fig.4.
The density dependence of the effective mass
$m^*$ (solution of Eqs.~(\ref{57})-(\ref{59})). The solid
line shows the results for symmetric matter ($\beta=0$). The
dashed and dotted lines show the proton and neutron
self-energies $m^{*(p)}$ and $m^{*(n)}$
in the neutron matter ($\beta=1$).

\noindent
Fig.5. The density dependence of  the functions
$x V_1$, $x V_2$, $x S_1$ and $x S_2$ introduced by
Eqs.~(\ref{112}), $x=\frac{\rho}{\rho_0}$. The dotted lines
demonstrate the quality of fitting with the simple functions on
$\rho$, as described in Appendix~D.

\noindent
Fig.6.
The density dependence of the single particle potential energy
$U$, Eq.~(\ref{24}. The solid line shows the results for
symmetric matter ($\beta=0$). The dashed and dotted lines show
the proton and neutron self-energies $U^{(p)}$ and $U^{(n)}$ in
the neutron matter ($\beta=1$).

\newpage

\begin{figure}
\centerline{\epsfig{file=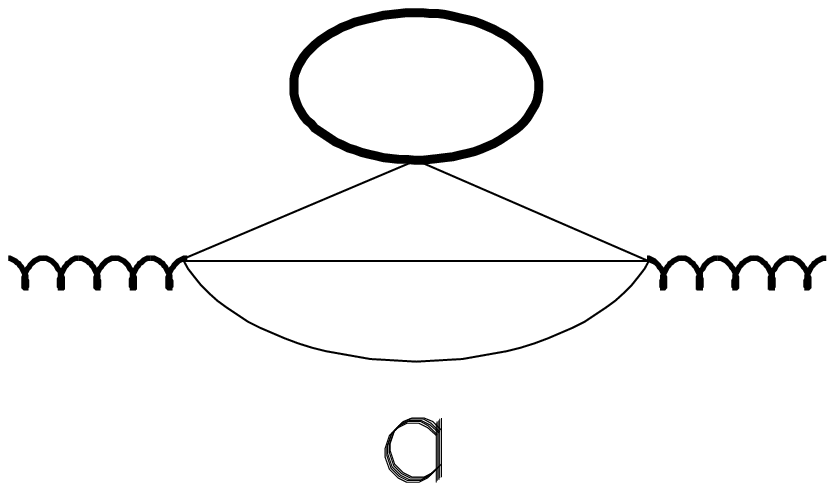,width=4cm}
\epsfig{file=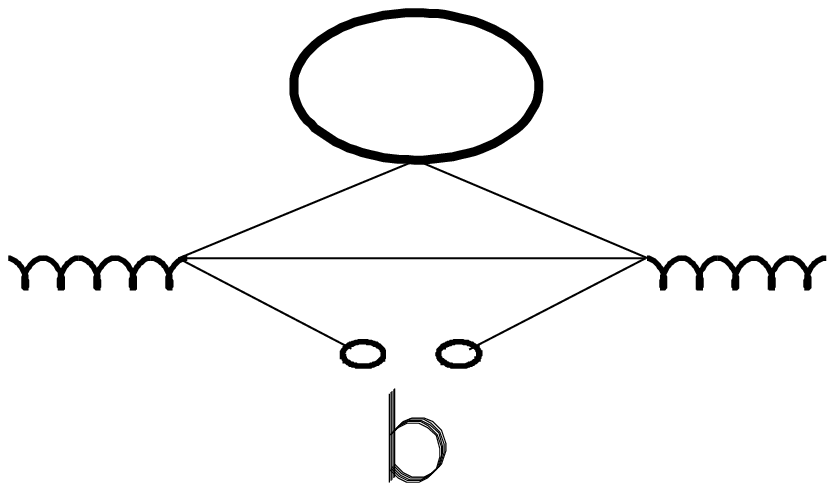,width=4cm}}
\vspace{-1cm}
\centerline{\epsfig{file=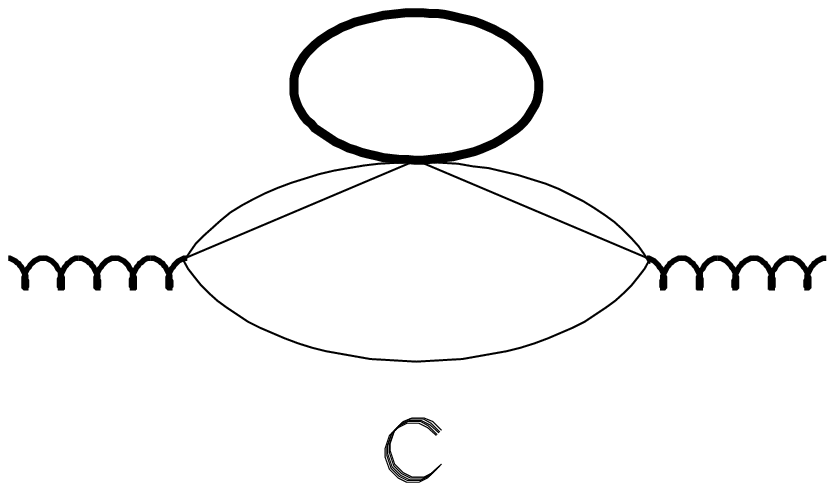,width=4cm}
\epsfig{file=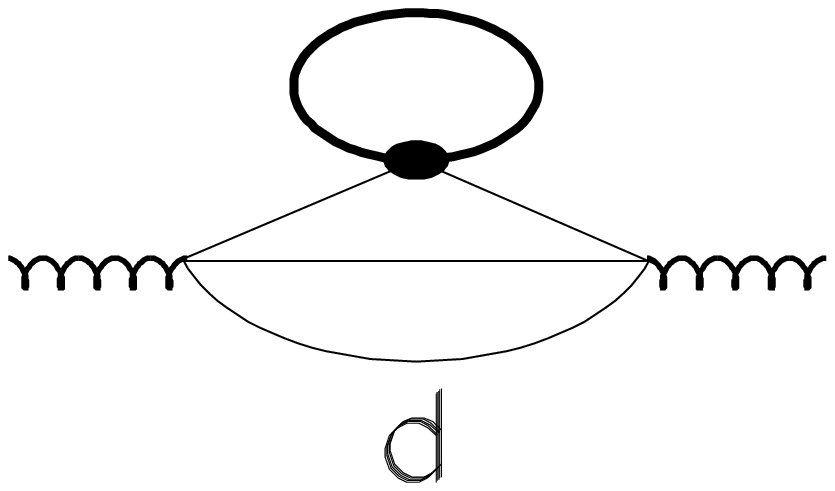,width=4cm}}
\caption{}
\end{figure}

\begin{figure}
\centerline{\epsfig{file=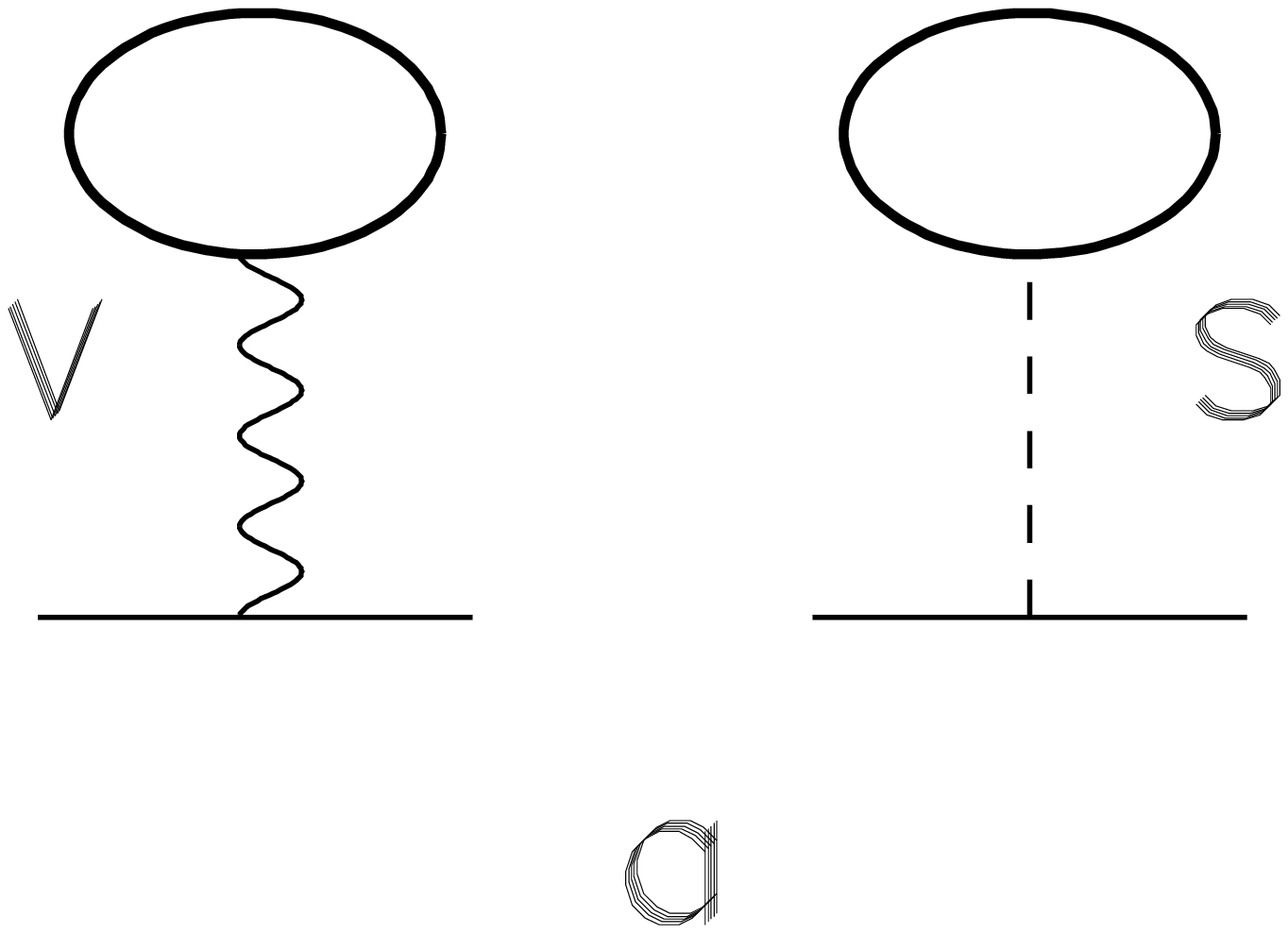,width=5cm}
\epsfig{file=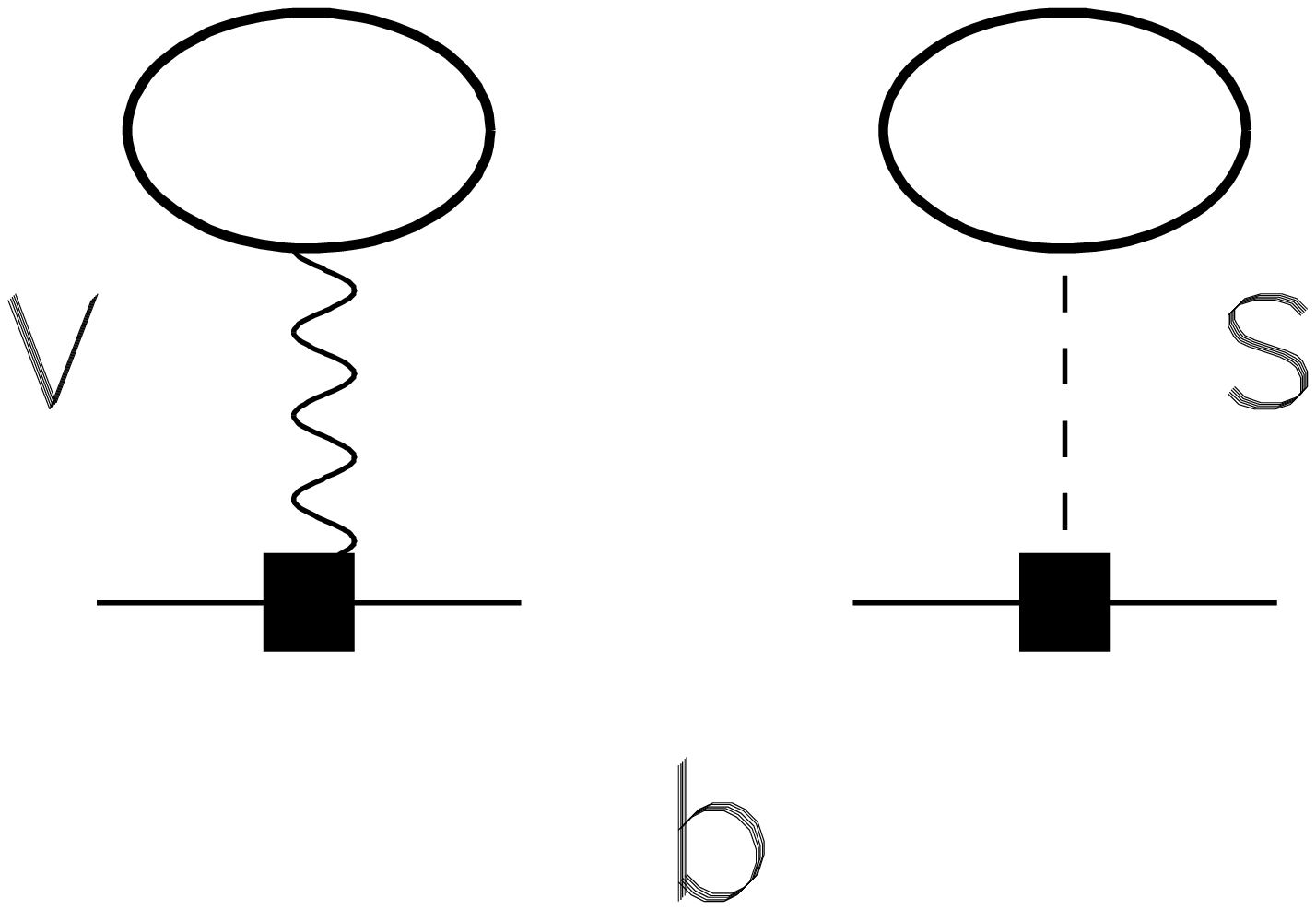,width=5cm}}
\centerline{\epsfig{file=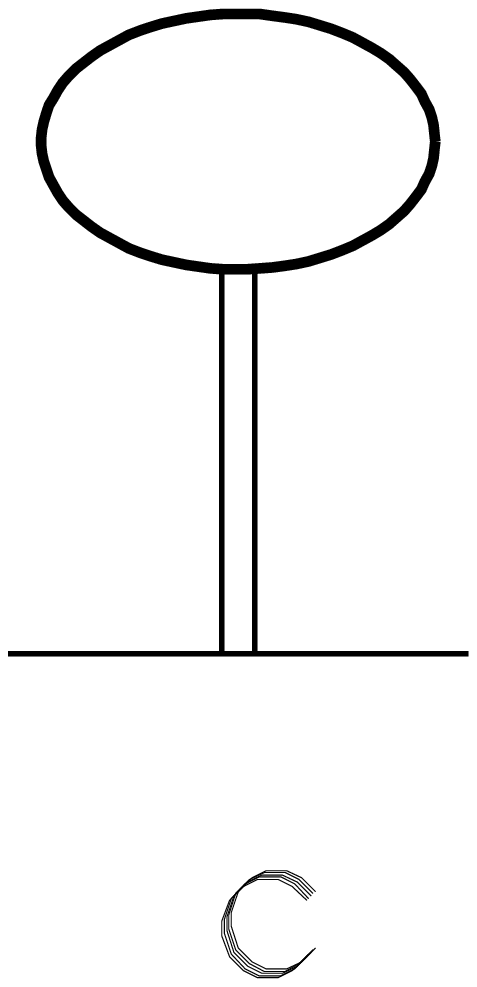,width=5cm}
\epsfig{file=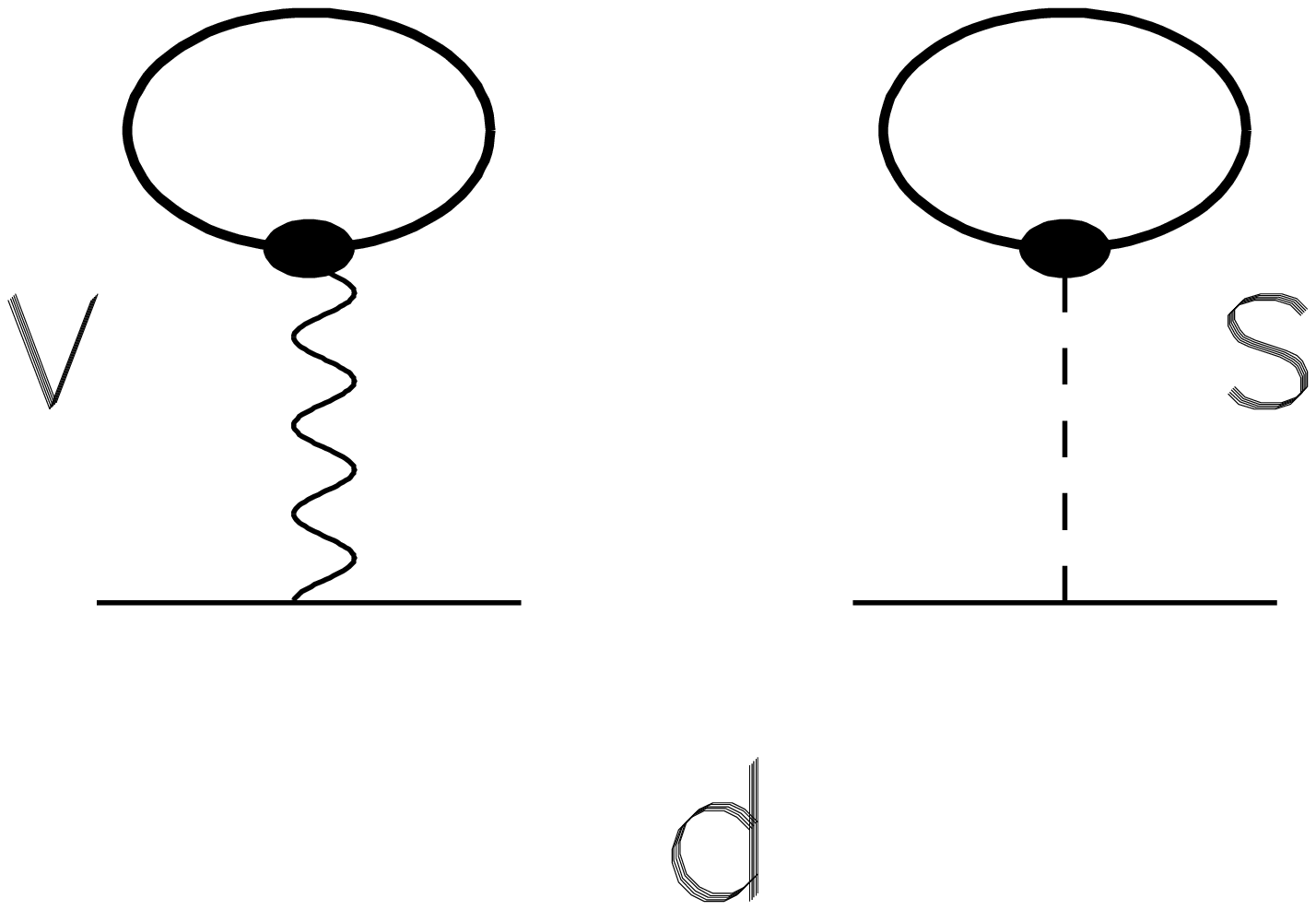,width=5cm}}
\caption{}
\end{figure}

\begin{figure}
\centering{\epsfig{figure=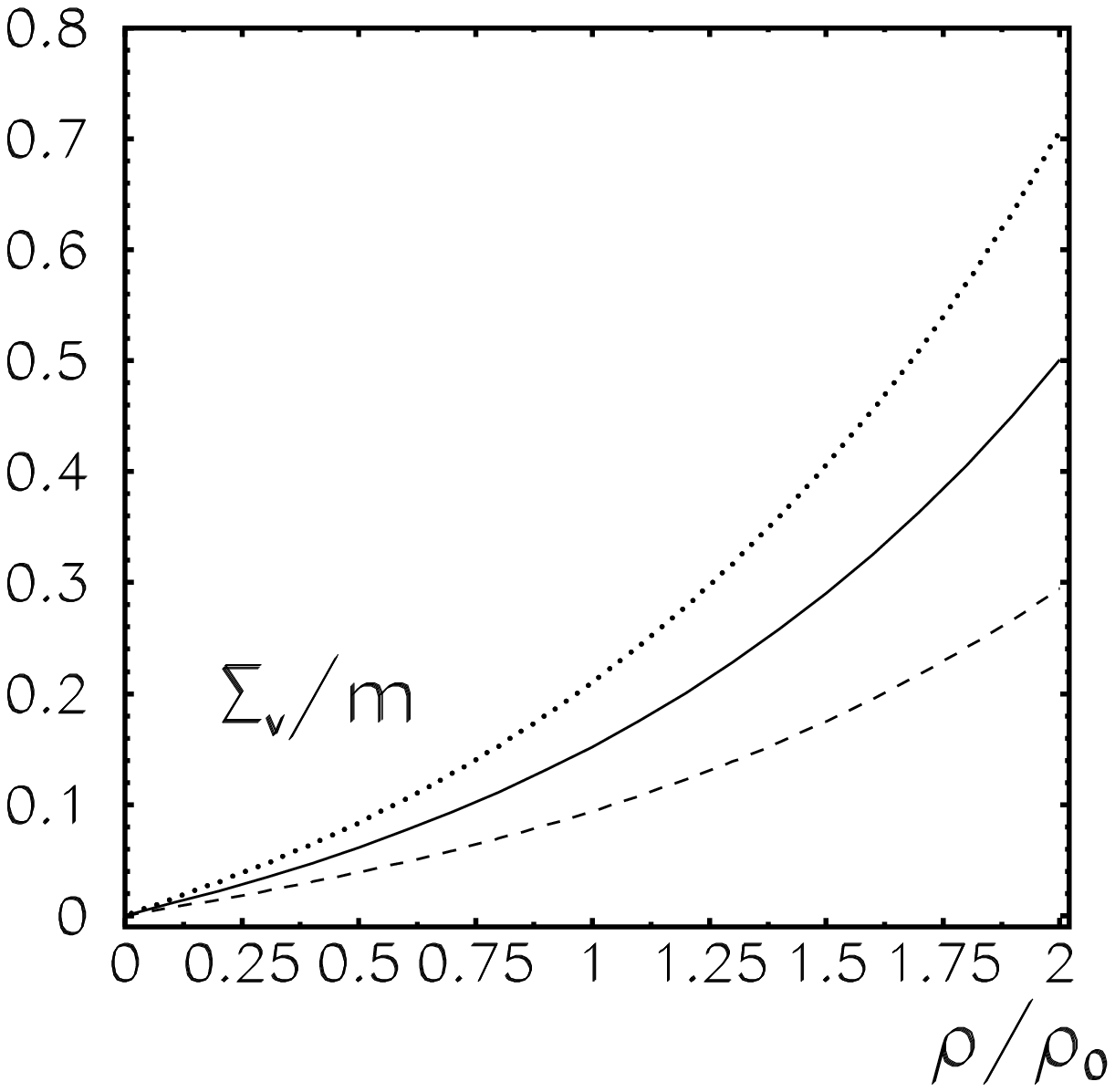,width=10cm}}
\caption{}
\end{figure}

\begin{figure}
\centering{\epsfig{figure=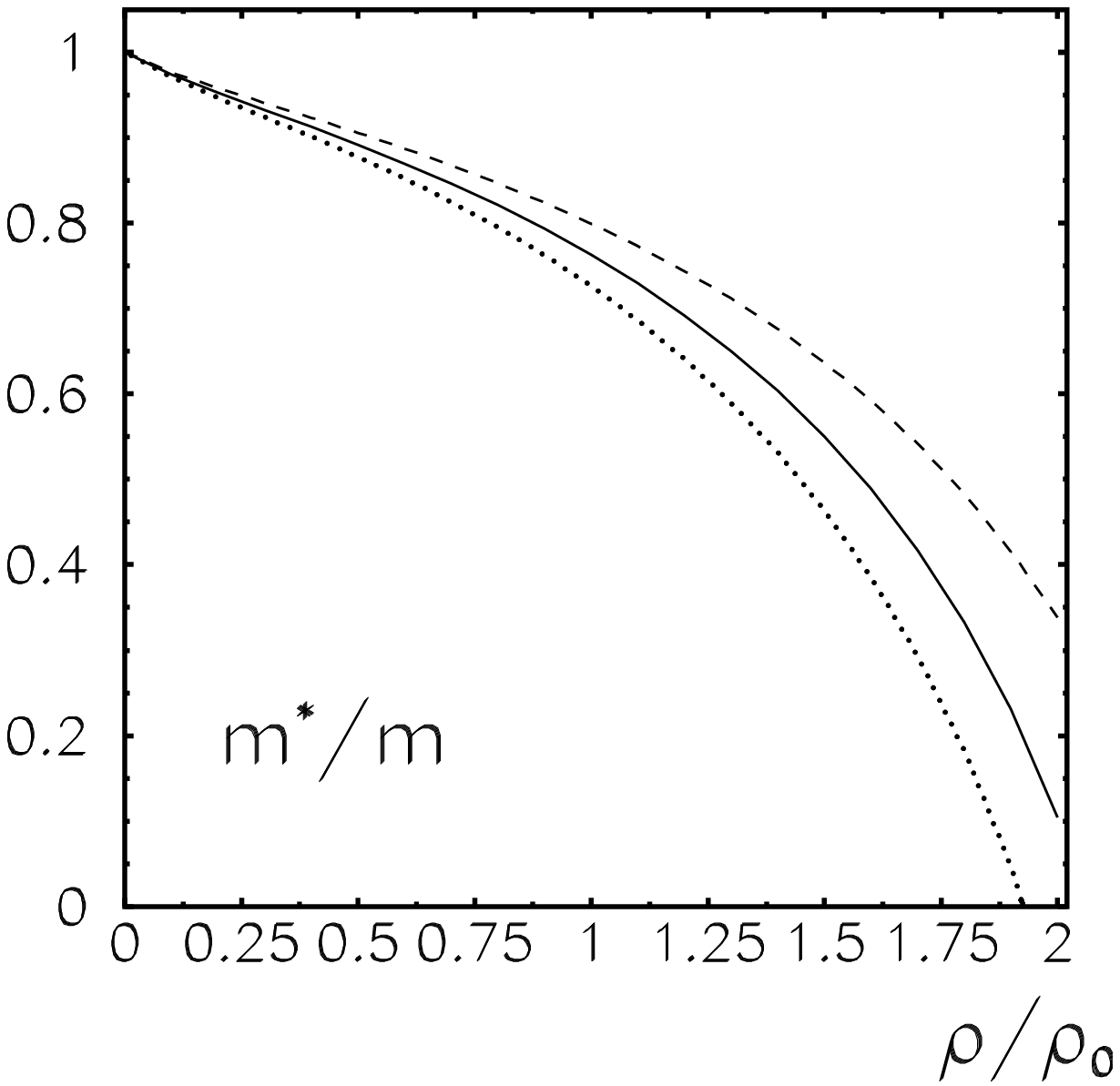,width=10cm}}
\caption{}
\end{figure}

\begin{figure}
\centering{\epsfig{figure=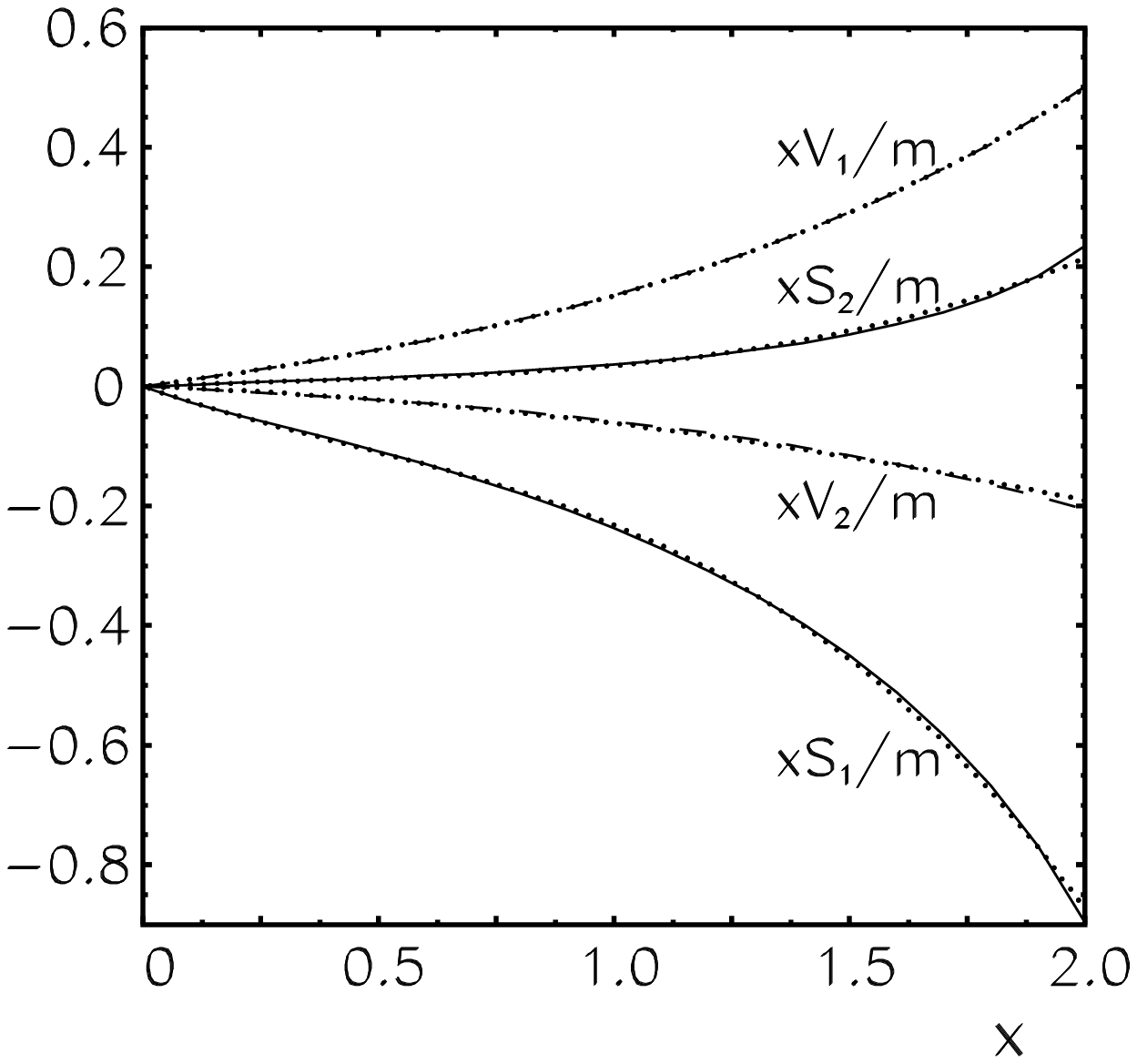,width=10cm}}
\caption{}
\end{figure}

\begin{figure}
\centering{\epsfig{figure=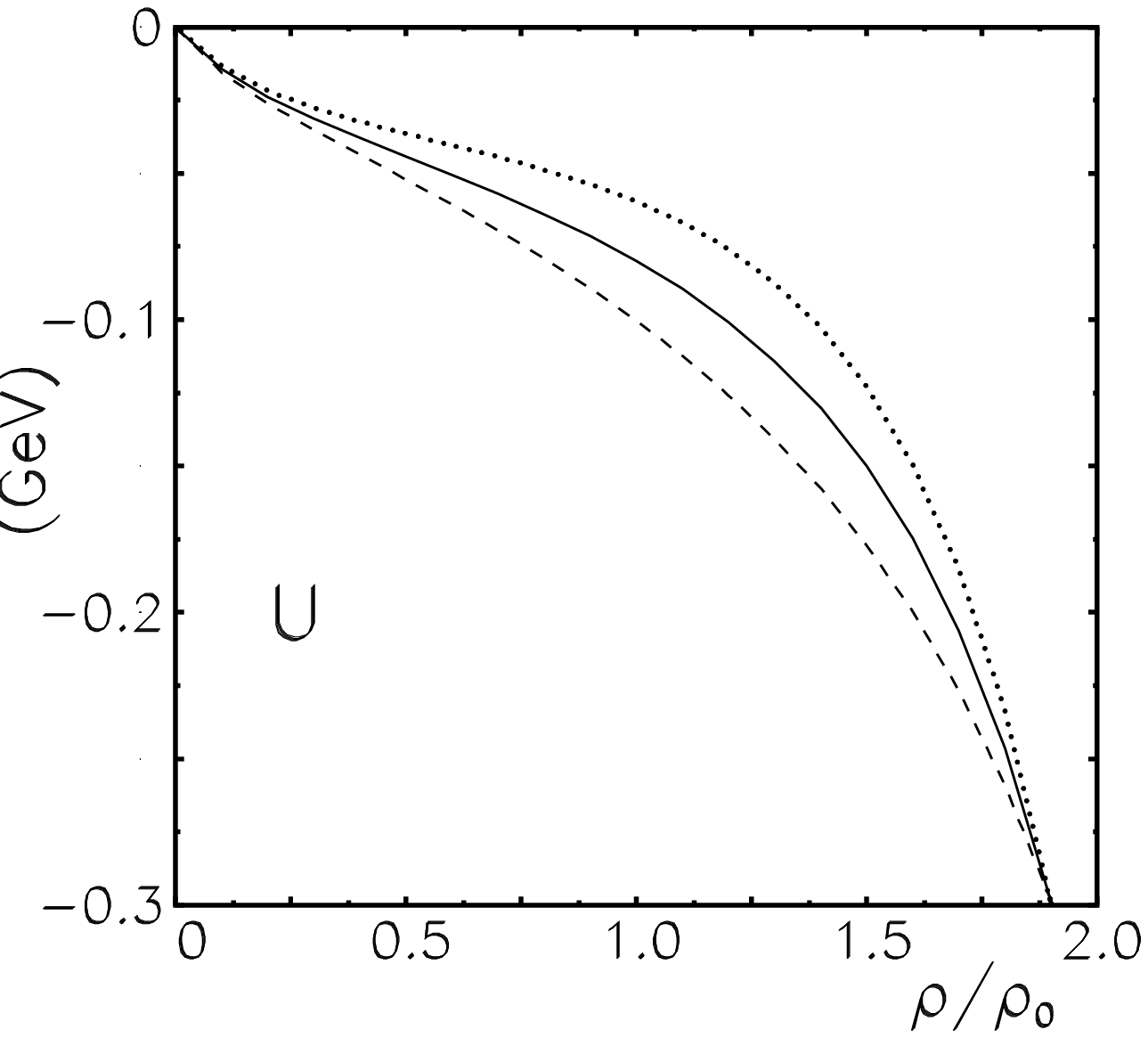,width=10cm}}
\caption{}
\end{figure}

 \end{document}